\documentclass{aa}
\usepackage[varg]{txfonts}
\usepackage{lipsum}  
\usepackage{graphicx}
\usepackage{natbib}
\usepackage{hyperref}
\bibpunct{(}{)}{;}{a}{}{,} 

\begin{document}
\title{Precise radial velocities of giant stars}
\subtitle{XVI. Planet occurrence rates from the combined analysis \\
	of the Lick, EXPRESS, and PPPS giant star surveys}
\author{Vera Wolthoff\inst{1}
 \and Sabine Reffert\inst{1}
 \and Andreas Quirrenbach\inst{1}
 \and Matías I. Jones\inst{2}
 \and Robert A. Wittenmyer\inst{3}
 \and James S. Jenkins\inst{4,5}}
\institute{Landessternwarte, Zentrum für Astronomie der Universit\"at Heidelberg, K\"onigstuhl 12, 69117 Heidelberg, Germany\\
 \email{vwolthoff@lsw.uni-heidelberg.de}
 \and European Southern Observatory, Alonso de Córdova 3107, Vitacura, Casilla 19001, Santiago, Chile
 \and University of Southern Queensland, Centre for Astrophysics, West Street, Toowoomba, QLD 4350 Australia
 \and Departamento de Astronomía, Universidad de Chile, Camino El Observatorio 1515, Las Condes, Santiago, Chile
 \and Centro de Astrofísica y Tecnologías Afines (CATA), Casilla 36-D, Santiago, Chile}
\date{Received 21 October 2021 / Accepted 11 February 2022}

\abstract {Radial velocity surveys of evolved stars allow us to probe a higher stellar mass range, on average, compared to main-sequence samples. Hence, differences between the planet populations around the two target classes can be caused by either the differing stellar mass or stellar evolution. To properly disentangle the effects of both variables, it is important to characterize the planet population around giant stars as accurately as possible.} {Our goal is to investigate the giant planet occurrence rate around evolved stars and determine its dependence on stellar mass, metallicity, and orbital period.} {We combine data from three different radial velocity surveys targeting giant stars: the Lick giant star survey, the radial velocity program EXoPlanets aRound Evolved StarS (EXPRESS), and the Pan-Pacific Planet Search (PPPS), yielding a sample of 482 stars and 37 planets. We homogeneously rederived the stellar parameters of all targets and accounted for varying observational coverage, precision and stellar noise properties by computing a detection probability map for each star via injection and retrieval of synthetic planetary signals. We then computed giant planet occurrence rates as a function of period, stellar mass, and metallicity, corrected for incompleteness.} {Our findings agree with previous studies that found a positive planet-metallicity correlation for evolved stars and identified a peak in the giant planet occurrence rate as a function of stellar mass, but our results place it at a slightly smaller mass of $(1.68\pm~0.59)\,M_\sun$. The period dependence of the giant planet occurrence rate seems to follow a broken power-law or log-normal distribution peaking at $(718\pm 226)$\,days or $(797\pm 455)$\,days, respectively, which roughly corresponds to 1.6 AU for a $1\,M_\sun$ star and 2.0 AU for a $2\,M_\sun$ star. This peak could be a remnant from halted migration around intermediate-mass stars, caused by stellar evolution, or an artifact from contamination by false positives. The completeness-corrected global occurrence rate of giant planetary systems around evolved stars is $10.7\%^{+2.2\%}_{-1.6\%}$ for the entire sample, while the evolutionary subsets of RGB and HB stars exhibit $14.2\%^{+4.1\%}_{-2.7\%}$ and $6.6\%^{+2.0\%}_{-1.3\%}$, respectively. However, both subsets have different stellar mass distributions and we demonstrate that the stellar mass dependence of the occurrence rate suffices to explain the apparent change of occurrence with the evolutionary stage.} {}
\titlerunning{Precise radial velocities of giant stars. XIII.}
\authorrunning{V. Wolthoff et al.} 
\maketitle

\section{Introduction} \label{sec_intro}
\begin{figure*}
	\sidecaption
	\includegraphics[width=12cm]{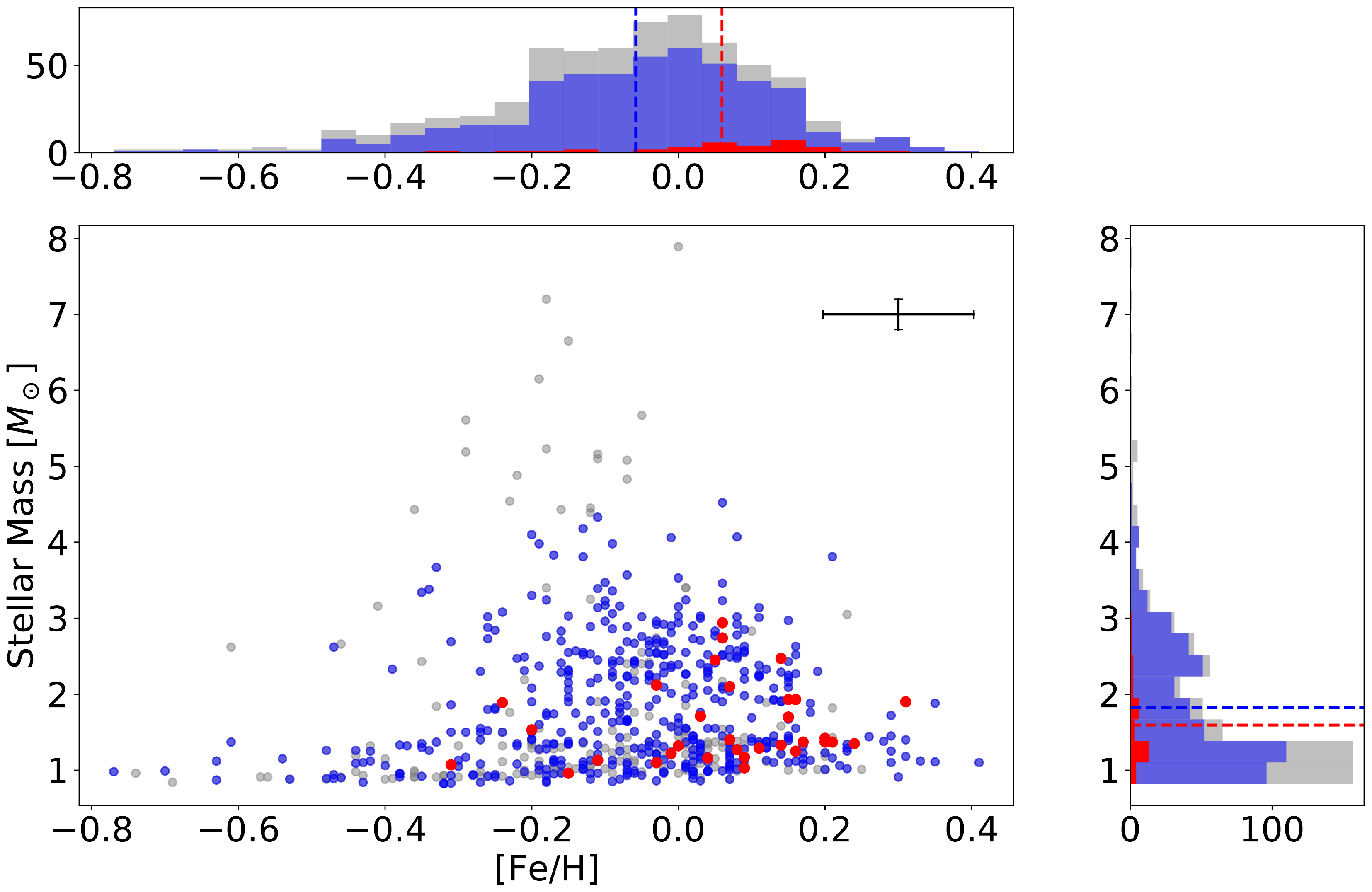}
	\caption{Stellar mass versus metallicity for the survey stars and histograms of both parameters. The complete sample is shown in gray; the 482 stars left after the homogenizing cuts (described in Sec.~\ref{sec_sample}) are shown in blue, and the red points identify the 32 planet hosting stars within the sample. The mean uncertainties are shown in the upper right corner. The red and blue dotted lines indicate the mean stellar mass and metallicity for the targets with and without planets, respectively, for the following: planet hosts, $\overline{M_*}=(1.59\pm0.51)~M_\sun$ and $\overline{[\mathrm{Fe/H}]}=(0.06\pm0.14)~\mathrm{dex}$; and stars without planets, $\overline{M_*}=(1.83\pm0.79)~M_\sun$ and $\overline{[\mathrm{Fe/H}]}=(-0.06\pm0.18)~\mathrm{dex}$.}
	\label{fig_mfeh}
\end{figure*}
To date, more than 4000 extrasolar planets have been discovered and several thousand candidates are still awaiting confirmation. And while they cover a large range of orbital parameters and are hosted by stars of many different types, a large portion of parameter space has yet to be explored in more depth.\par
Most of the known planet hosts have masses comparable to or smaller than that of the Sun because they are excellent targets for both radial-velocity (RV) and transit planet searches. In contrast, stars more massive than $\sim$$\,1.5M_\sun$ are difficult targets for most of their lives: their size impedes detecting planetary transits, and the achievable RV precision is reduced because, compared to their lower-mass counterparts, A-F stars only show few absorption lines in their spectra (due to their higher surface temperatures) and they rotate faster leading to rotational broadening of the lines \citep{lagrange2009}. But luckily, when these stars leave the main sequence, they cool down and rotate more slowly, becoming feasible targets for RV planet searches. Thus, evolved stars have been targeted by several RV surveys: the Lick giant star survey \citep{frink2001, reffert2015}, the Okayama Planet Search Program \citep{sato2005}, the ESO planet search program \citep{setiawan2003}, the Tautenburg Observatory Planet Search \citep{hatzes2005}, the survey ``Retired A Stars and Their Companions'' \citep{johnson2007}, the Penn State-Toru{\'n} Planet Search \citep{niedzielski2007} including its follow-up program ``Tracking Advanced Planetary Systems’’ \citep{niedzielski2015}, the BOAO K-giant survey \citep{han2010}, the Pan-Pacific Planet Search \citep[PPPS, ][]{wittenmyer2011b}, and the EXoPlanet aRound Evolved StarS project \citep[EXPRESS, ][]{jones2011}. To date, more than 150 planets around evolved hosts have been discovered.\par
Addressing the dependence of planet occurrence on stellar mass, \citet{johnson2010} studied a sample of main-sequence and subgiant stars with masses ranging from 0.2 to $2.0\,M_\sun$ and found an approximately linear increase in the giant planet occurrence rate with stellar mass. These results were updated by \citet{ghezzi2018}. However, beyond $\sim2\,M_\sun$, the occurrence rate appears to quickly decrease again \citep{reffert2015, jones2016}. Moreover, while several earlier studies found no evidence for a dependence of planet occurrence around giant stars on stellar metallicity \citep{pasquini2007,takeda2008,maldonado2013}, other studies found that the well-known planet-metallicity correlation \citep{fischervalenti2005}, which states that giant planets are preferentially found around metal-rich host stars, also seems to hold for evolved host stars \citep{hekker2007,johnson2010, reffert2015, jones2016, wittenmyer2017a}.\par
Considering the properties of the planets hosted by post-main sequence stars, \citet{bowler2010} found that the planet mass-period distribution is significantly different for their sample of subgiant hosts compared to the one derived for main-sequence stars \citep{cumming2008}, while the orbital eccentricities are smaller for planets around evolved stars \citep{jones2014,maldonado2013}. The most notable difference is that close-in giant planets (Hot and Warm Jupiters with $P$\,$<$\,$90\,\mathrm{days}$) are very rare \citep{johnson2007,sato2008}. While it was theorized that formation of these planets might be hindered for higher stellar masses due to faster disk dispersal and longer migration timescales \citep{kennedy2008a, kennedy2009, currie2009}, it now becomes apparent that Hot Jupiters do exist around A stars \citep[see e.g., ][]{sivert2018,hellier2019,rodriguez2020}. Based on transit surveys, \citet{zhou2019b} find a Hot Jupiter occurrence rate of $[0.26\pm 0.11]\% $ for A stars, and compared to their values for G stars ($[0.71\pm 0.31]\%$) and F stars ($[0.43\pm 0.15]\%$), there seems to be a slight trend with stellar mass. In particular, the values for G and A stars differ at the $1\sigma$ level.\par
Hence, the observed paucity may be caused by stellar evolution which could have had an important impact on the architecture of the observed planetary systems, ultimately removing close-in planets by engulfment during expansion on the giant branch \citep{villaver2009}. Planet engulfment has also been discussed as a possible explanation for the unusually high Lithium abundances observed in a small fraction of giant stars \citep[see, e.g., ][]{adamow2012}. However, stellar tides and mass loss are weaker for stars not undergoing a Helium-flash ($M_*$\,$\gtrsim$\,$2\,M_\sun$) and may not be efficient at removing Warm Jupiters \citep{kunitomo2011, villaver2014}; however, many uncertainties remain in the evolutionary simulations, and accurately characterizing the planet population around post-main sequence stars (and their progenitors of different masses) can be helpful to constrain the effect of stellar evolution on planetary orbits and theoretical models of planet engulfment.\par
In this study, we combine three of the previously mentioned RV surveys (the Lick giant star survey, EXPRESS, and PPPS) to investigate the planet population around evolved stars and its correlations with host star properties. In Sec.~\ref{sec_obs}, we describe the observations of the three surveys. Section~\ref{sec_sample} presents the stellar sample, while we provide an overview of the detected substellar companions in Sec.~\ref{sec_planets}. In Sec.~\ref{sec_detprobs}, we summarize the derivation of detection probabilities for the combined sample. The global planet occurrence rate is computed in Sec.~\ref{sec_occ_global}. Section~\ref{sec_occ_period} looks at the planet period distribution, while Sec.~\ref{sec_occ_mass} analyzes the functional dependence of planet occurrence rate on stellar mass and metallicity. We finally conclude in Sec.~\ref{sec_discussion}.\par

\section{Observational data} \label{sec_obs}
For our analysis, we combine observations from three RV surveys: the Lick giant star survey \citep{reffert2006}, the EXoPlanets aRound Evolved StarS (EXPRESS) survey \citep{jones2011}, and the Pan-Pacific Planet Search \citep[PPPS,][]{wittenmyer2011b}.
\subsection{Lick observations}
The Lick giant star survey targets 372 bright ($V\le 6\,\mathrm{mag}$) G~and K~giants. The survey started in 1999 with an initial sample size of 86 K giants with small photometric variability. One year later, 96 stars were added to the sample with relaxed constraints on photometric variability and three stars were removed as they turned out to be visual binaries. In 2004, another 194 G and K giant stars with on average higher masses and bluer colors ($0.8\le B-V\le 1.2$) joined the sample. More details on the selection criteria are outlined in \citet{frink2001} and \citet{reffert2015}. The stars were observed at UCO/Lick Observatory on Mount Hamilton in California using the 0.6\,m Coudé Auxiliary Telescope (CAT) with the Hamilton High Resolution Echelle Spectrograph ($R\sim60\,000$). RVs were extracted using the iodine cell technique \citep{butler1996}, achieving a typical radial velocity precision of 5\,$\mathrm{m\,s^{-1}}$ to 8\,$\mathrm{m\,s^{-1}}$. Observations came to a preliminary halt in 2011 when the iodine cell at Lick broke; but they will be resumed in the near future with a new high resolution echelle spectrograph built for the 72\,cm Waltz Telescope located at the Landessternwarte Königstuhl in Heidelberg \citep{tala2016}. In the mean time, additional observations of the most interesting targets were made with the 1m Hertzsprung SONG telescope \citep{grundahl2007,grundahl2017} on Tenerife (which are however not included in this analysis in order to keep the data more homogeneous).
\subsection{EXPRESS observations}
The EXPRESS survey began observing a sample of 164 bright ($V\le 8\mathrm{mag}$) giant stars in 2009 -- for details on sample selection see \citet{jones2011}. Three different spectrographs were used within the survey: Early observations were made with the fiber echelle spectrograph (FECH; $R\sim45\,000$) at the 1.5\,m telescope at the Cerro Tololo Inter-American Observatory (CTIO), which however only reach a long-term precision of $\sim$20$-$30\,$\mathrm{m\,s^{-1}}$ and are not considered in our analysis. In 2011, FECH was replaced by CHIRON \citep{tokovinin2013}, which in turn is able to provide a resolution of  $R\sim90\,000$ and a mean RV precision of $\sim$5\,$\mathrm{m\,s^{-1}}$ and has been used for the EXPRESS observations since. An iodine cell is used for wavelength reference. Additional data have been obtained since 2010 with the Fiber-fed Extended Range Optical Spectrograph \citep[FEROS; ][]{kaufer1999} on the 2.2\,m telescope at La Silla Observatory, which reaches a resolution of $R\sim48\,000$. RVs were computed from the FEROS spectra using the cross-correlation technique and the simultaneous calibration method \citep{baranne1996}; more details on the data reduction are given in \citet{jones2017}. The FEROS data likewise reach a long-term precision of~$\sim$5\,$\mathrm{m\,s^{-1}}$.
\subsection{PPPS observations}
The PPPS target list includes 164 evolved stars which were selected as a Southern hemisphere extension to the Lick \& Keck Observatory survey of ``retired A stars'' \citep[e.g.,][]{johnson2010}; for details see \citet{wittenmyer2011b, wittenmyer2016c}. Observations were taken between 2009 and 2015 using the University College London Echelle Spectrograph \citep[UCLES; ][]{diego1990} at the 3.9\,m Anglo-Australian Telescope (AAT), which achieves a resolution of $R\sim45\,000$. Precise RVs are determined by the iodine-cell technique \citep{butler1996}. While the survey as a whole has been discontinued, selected targets showing promising candidate planet signals are continued to be monitored as a ``PPPS Legacy'' program with the Minerva-Australis telescope array and with the SONG telescopes at Tenerife and Australia \citep{wittenmyer2018, addison2019}.

\section{Stellar sample} \label{sec_sample}
\begin{figure}
	\resizebox{\hsize}{!}{\includegraphics{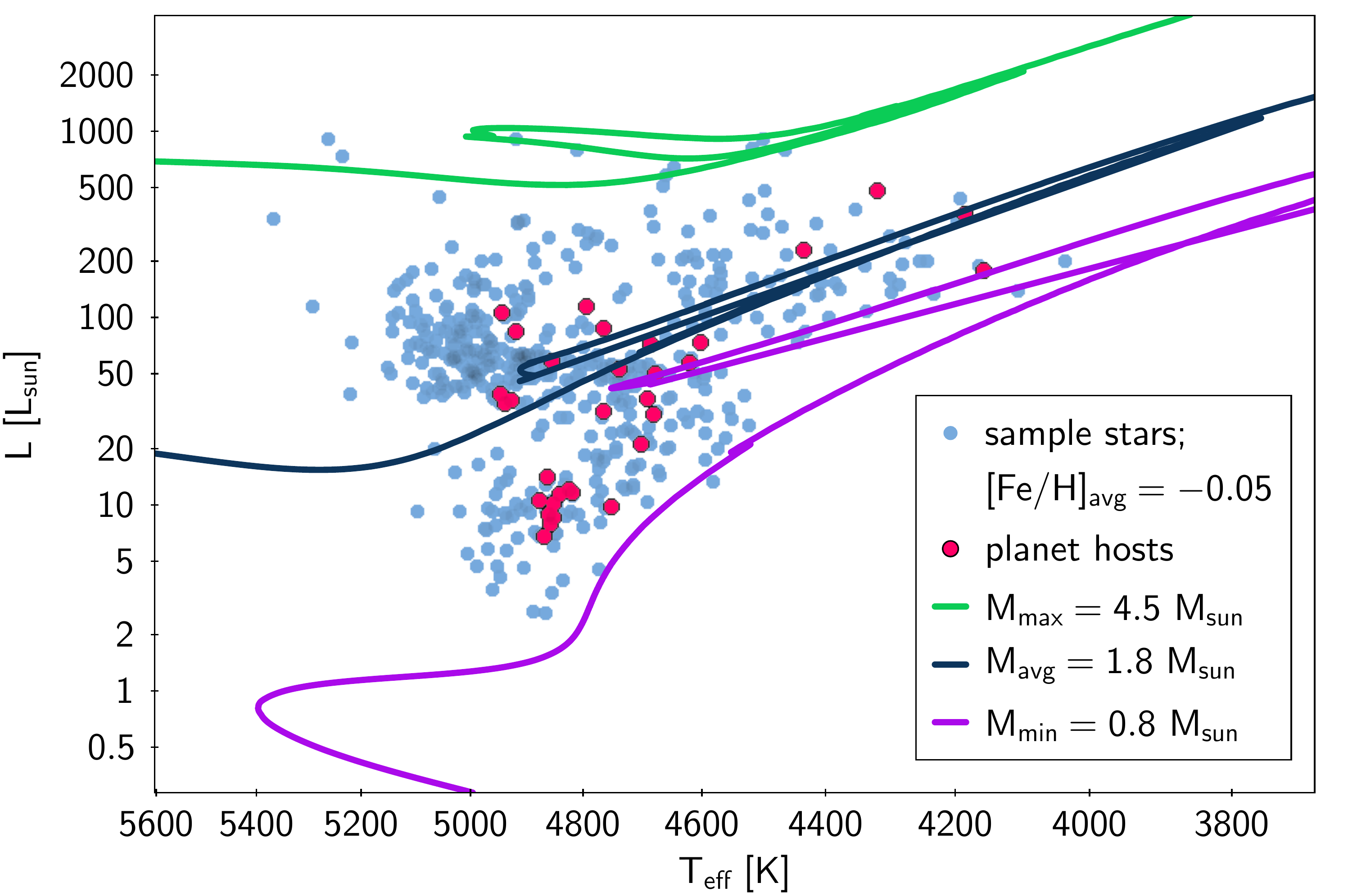}}
	\caption{Hertzsprung-Russell diagram of the sample stars after cuts (blue). Planet hosting stars are shown in red. The green, dark-blue, and violet lines correspond to the PARSEC evolutionary tracks for the maximum ($4.5\,M_\sun$), mean ($1.8\,M_\sun$), and minimum mass  ($0.8\,M_\sun$) among the target stars, respectively, and a metallicity of [Fe/H]=$-0.05$, which is the sample mean.}
	\label{fig_hrd}
\end{figure}

\begin{figure*}
	\centering
	\includegraphics[width=6.5cm]{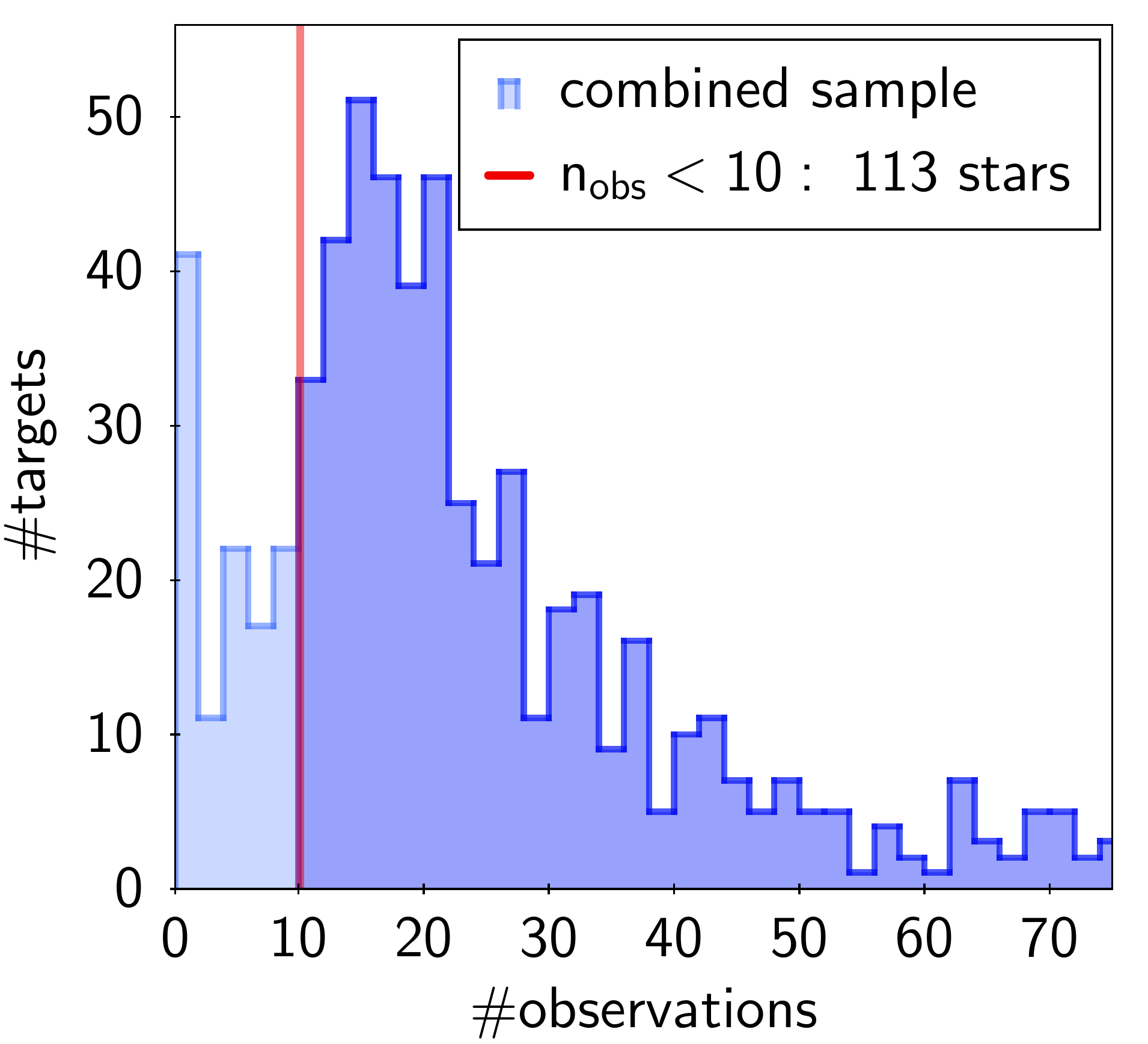}
	\includegraphics[width=6.5cm]{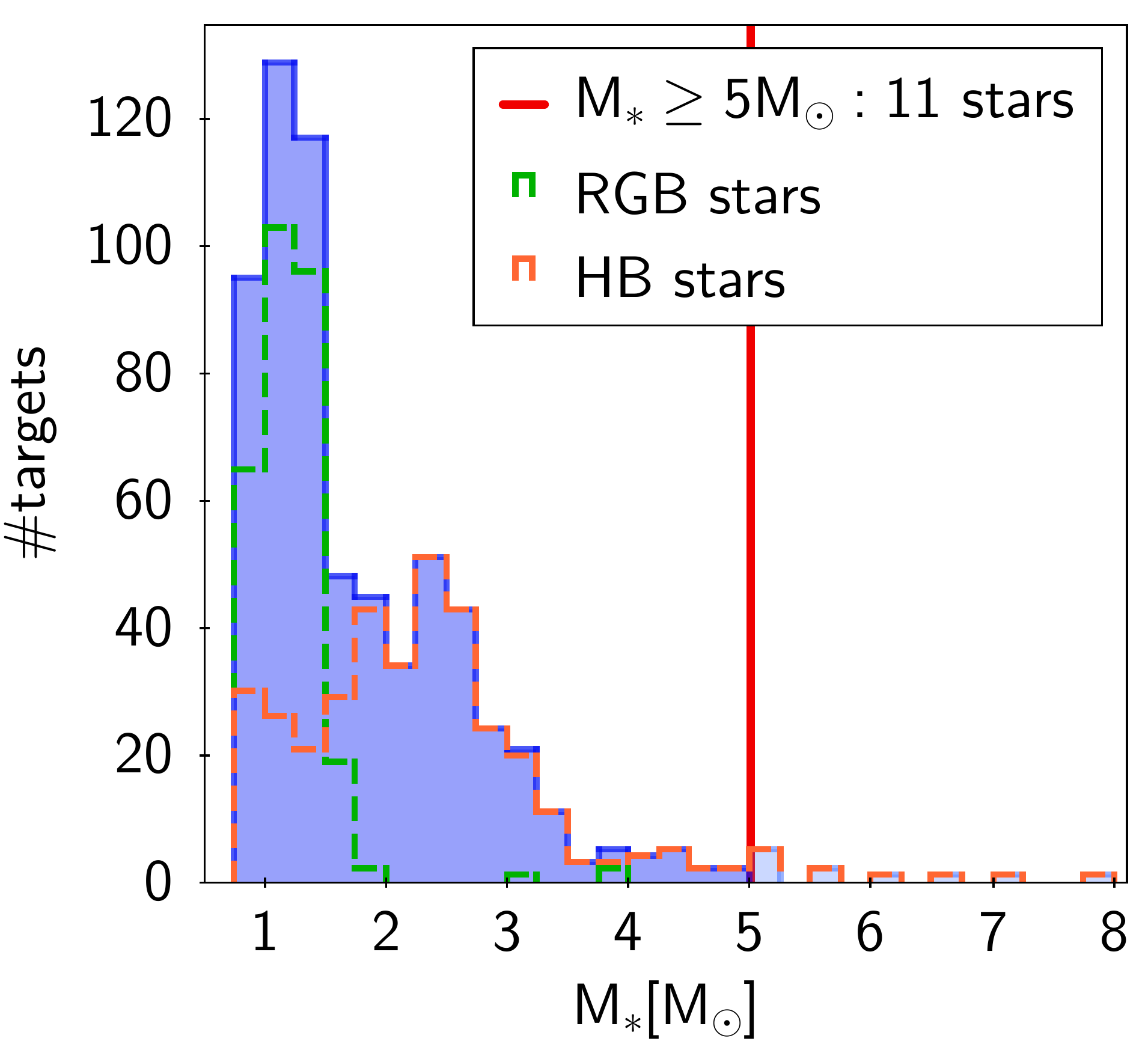}
	\includegraphics[width=6.5cm]{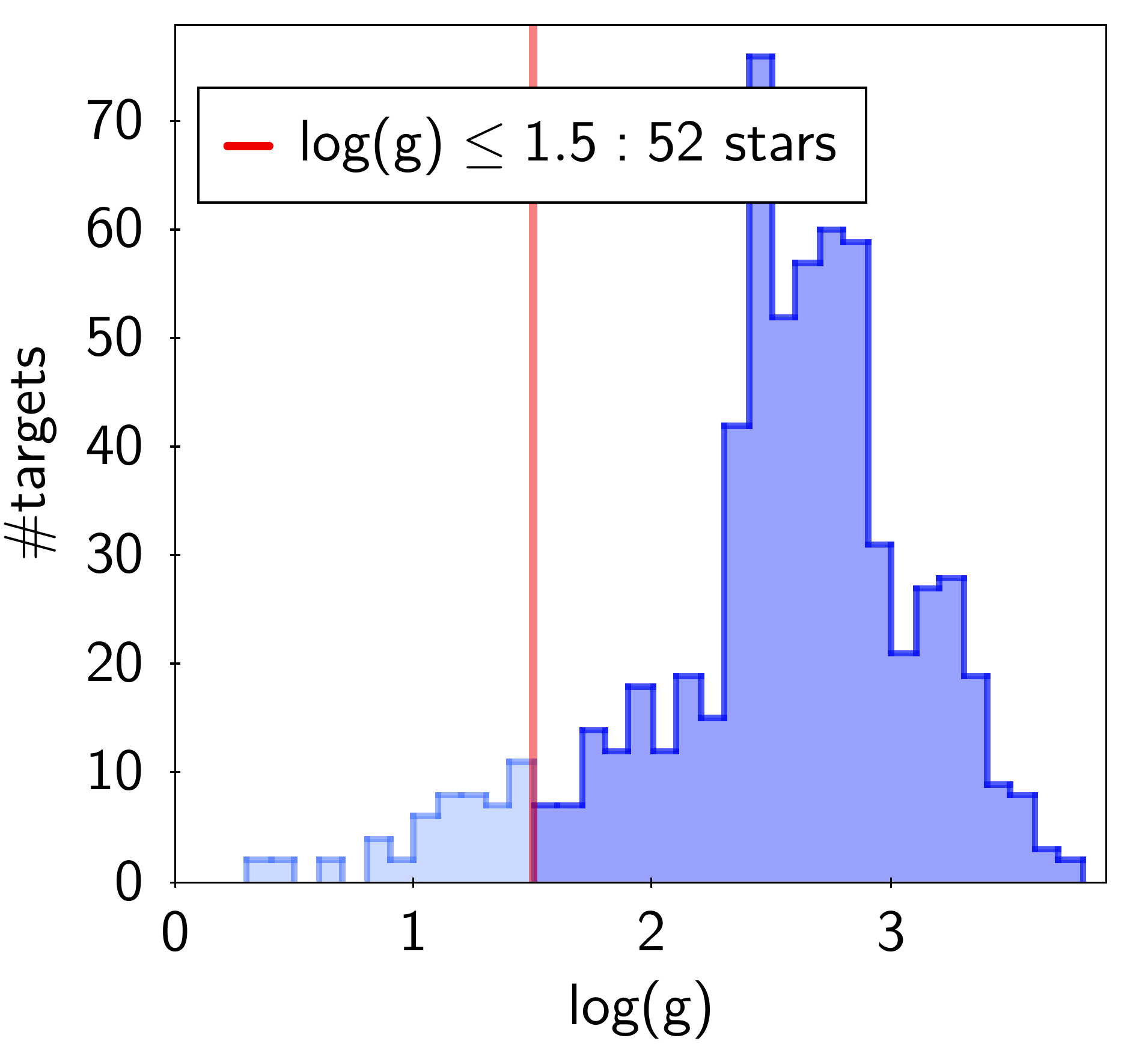}				
	\includegraphics[width=6.5cm]{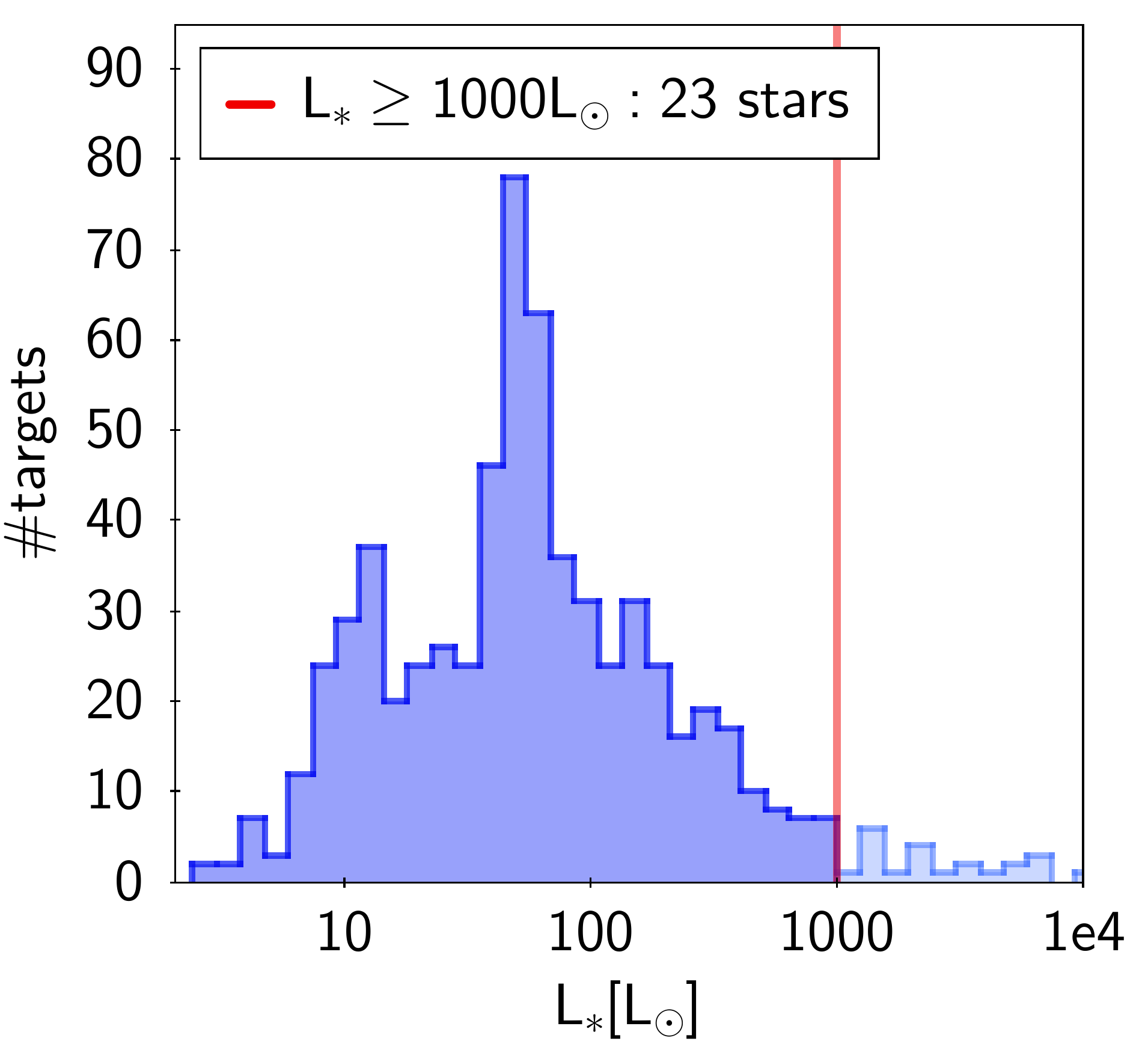}
	\caption{Histograms to illustrate the homogenizing cuts applied to the target stars. \textit{Top left}: Distribution of the number of radial velocity data points, stars with 10 observations or less are excluded (the plot is truncated toward larger numbers; more than 50 stars have more than 70 RV data points). \textit{Top right}: Distribution of the stellar mass; stars as massive as 5~$M_\sun$ or more are excluded. The mass distributions of the stars more likely to be on the RGB and HB are shown in green and orange, respectively. \textit{Bottom left}: Distribution of stellar surface gravity; stars with $\log(g)$ equal to or less than 1.5 are excluded. \textit{Bottom right}: Distribution of stellar luminosity; stars as luminous as $1000~L_\sun$ or more are excluded.}
	\label{fig_cuts}
\end{figure*}

Accounting for partial overlaps between the target lists of the three surveys (the EXPRESS and PPPS surveys share 37 stars, Lick and EXPRESS share 12 stars, and the PPPS and Lick target lists have one star in common), the combined sample comprises 650 stars. We homogeneously rederive the stellar parameters (mass, radius, age, surface gravity, effective temperature, and luminosity) using the method presented in \citet{stock2018}. A Python implementation of the original, unpublished IDL code applying the method of \cite{stock2018}, which is titled \texttt{SPOG} (\textbf{S}tellar \textbf{P}arameters \textbf{o}f \textbf{G}iants), was recently made available by the author on github\footnote{\url{https://github.com/StephanStock/SPOG}}. It relies on Bayesian inference to compare spectroscopic, photometric, and astrometric observables to grids of PARSEC evolutionary tracks \citep{bressan2012} while incorporating knowledge about evolutionary timescale and initial mass function as prior information into the analysis. There has been some controversy in the literature regarding the parameter determinations of evolved stars in RV planet searches. Criticism was raised that the stellar masses might be systematically overestimated \citep{lloyd2011, lloyd2013, schlaufman2013}. Several other studies rebutted this, finding no evidence for a considerable systematic offset \citep{johnson2013, ghezzi2015, north2017, ghezzi2018}. Indeed it seems that spectroscopic masses tend to be overestimated if not accounting for the evolutionary timescales in different regions of the Hertzsprung-Russell-Diagram. \citet{stock2018} demonstrate the reliability of their method (which takes into account evolutionary timescales and the initial mass function, and fits the red-giant branch (RGB) and the horizontal branch (HB) separately) by applying the method to a test sample of 26 giant stars with available astroseismic mass determinations and find no indications for a significant overestimation in the determined stellar masses. The reliability is also confirmed by \citet{malla2020}, who note that the method by \citet{stock2018} shows the smallest mass offset among their comparison of spectroscopic to asteroseismic mass estimates. We are therefore confident that the rederived parameters are suitable for determining the dependence of planet occurrence rate on stellar mass.\par 
Recently, \citet{soto2020arXiv} reanalyzed the stellar properties of the EXPRESS sample, incorporating a reselection of iron lines, newer extinction maps, and an improved interpolation between evolutionary tracks. For a subset of stars with available astroseismic parameters, they find an excellent agreement between their method and asteroseismology. Overall, the masses of the EXPRESS target stars are found to be on average $0.28\, M_\sun$ smaller than previously determined \citep{jones2011}. This is in good agreement with the results using the method from \citet{stock2018}: we find the mean difference between the old and new masses of the EXPRESS stars to be $-0.45\pm0.25\,M_\sun$, with the mean mass shifting from $1.86\pm0.52\,M_\sun$ to $1.41\pm0.48\,M_\sun$. Also the results for the evolutionary subsamples agree well: \citet{soto2020arXiv} find a mean mass of $1.41\pm0.46\,M_\sun$ and $1.87\pm0.53\,M_\sun$, while we find $1.21\pm0.17\,M_\sun$ and $1.74\pm0.61\,M_\sun$ for the RGB and HB stars, respectively.\par
The Bayesian inference method from \citet{stock2018} supplies two probability distribution functions for each of the stellar parameters listed above, assuming the star is either ascending the RGB or has reached the HB, together with a probability estimate for the correct assignment to one of these evolutionary stages. In our analysis, we use the mode values of the distributions for the more probable evolutionary stage. The method requires the stellar metallicities as input which have been determined spectroscopically: for Lick by \citet{hekker2007}, for EXPRESS by \citet{jones2011}, and for PPPS by \citet{wittenmyer2016c}. The distribution of stellar masses and metallicities of the sample stars is shown in Fig.~\ref{fig_mfeh} with stars which have been identified as planet hosting stars (see Sec.~\ref{sec_planets}) marked in red. Figure~\ref{fig_hrd} illustrates the positions of the sample stars in the Hertzsprung-Russell-Diagram with respect to three selected PARSEC evolutionary tracks. Again, the planet hosting stars are plotted in red.\par
In a further effort to obtain a more homogeneous sample for our analysis, we apply several cuts to the combined sample, which are presented in Fig.~\ref{fig_cuts}. We do not consider stars for our analysis that have a high stellar mass, $M_*\ge 5\,M_\sun$, have a high luminosity, $L_* \ge 10^3\,L_\sun$, are most likely supergiants, $\log(g) \le 1.5$, or have less than 10 observations, $\#\mathrm{RVs} < 10$. As very massive, very luminous, and supergiant stars can exhibit many (quasi-) periodic signals of stellar origin, the cuts also serve to exclude targets with a more complex RV behavior where planet detection is hampered from the outset and at the same time, the risk of false positives is larger. Overall, we are left with 482 stars for the detectability and occurrence rate analysis.

\section{Planet detections in the combined sample} \label{sec_planets}
\begin{sidewaystable*}
	\caption{Planetary systems identified around stars from all three surveys.}
	\label{table_planets}
	\centering
	{\tiny
		\begin{tabular}{r c c c c c c c c c c c c l}
			\hline\hline\\ [-1.7ex]
			HIP&$M_*$&$\log~g$&$L_*$&[Fe/H]&evol.&$M_\mathrm{P}\sin(i)$&$P$&$K_\mathrm{RV}$&$a$&$e$&$f_\mathrm{R}$&$f_\mathrm{c}$&Detection reference\\
			&$[M_\sun]$&$[\mathrm{cm\ s^{-2}}]$&$[L_\sun]$&[dex]&stage&$[M_\mathrm{Jup}]$&$[\mathrm{d}]$&$[\mathrm{m\ s^{-1}}]$&$[\mathrm{AU}]$&&[\%]&[\%]&\\ [0.5ex] 
			\hline \\ [-1.7ex]   
			5364 & 1.4$\pm$0.2 & 2.3$\pm$0.1 &  74.2$\pm$$ 1.0$ &  0.07$\pm$0.10 & HB &  2.3$\pm$0.1 &  405.4$\pm$$  0.7$ &  49.8$\pm$$ 2.4$ & 1.2$\pm$0.1 & 0.06$\pm$0.03 & 100 &  66 & \citet{trifonov2014}\\
			&  &  &   &  & &  3.0$\pm$0.1 &  751.9$\pm$$  3.3$ &  54.1$\pm$$ 2.4$ & 1.8$\pm$0.1 & 0.14$\pm$0.05 & 100 &  66 & \citet{trifonov2014}\\
			8541 & 1.0$\pm$0.1 & 2.8$\pm$0.1 &  21.0$\pm$$ 2.9$ & -0.15$\pm$0.08 & RGB &  4.6$\pm$0.2 & 1563.7$\pm$$ 18.5$ &  83.8$\pm$$ 3.6$ & 2.6$\pm$0.1 & 0.13$\pm$0.04 & 100 &  72 & \citet{jones2016}\\
			16335 & 1.5$\pm$0.2 & 1.5$\pm$0.1 & 357.6$\pm$$15.2$ & -0.20$\pm$0.10 & RGB &  7.0$\pm$0.5 &  574.7$\pm$$  3.2$ & 129.8$\pm$$ 8.6$ & 1.6$\pm$0.1 & 0.18$\pm$0.07 & 100 &  94 & \citet{lee2014a}\\
			19011 & 2.1$\pm$0.6 & 2.5$\pm$0.1 &  88.0$\pm$$ 6.1$ & -0.03$\pm$0.10 & HB &  2.4$\pm$0.1 &  456.8$\pm$$  1.2$ &  38.8$\pm$$ 2.3$ & 1.5$\pm$0.2 & 0.09$\pm$0.05 & 100 &  68 & \citet{tala2020}\\
			20889 & 2.5$\pm$0.3 & 2.7$\pm$0.1 &  84.3$\pm$$ 1.6$ &  0.05$\pm$0.10 & HB &  7.0$\pm$0.2 &  581.8$\pm$$  2.1$ &  93.4$\pm$$ 2.4$ & 1.8$\pm$0.1 & 0.09$\pm$0.03 & 100 &  94 & \citet{sato2007}\\
			24275 & 1.4$\pm$0.1 & 3.2$\pm$0.1 &  11.3$\pm$$ 1.3$ &  0.17$\pm$0.10 & RGB &  1.7$\pm$0.2 &  555.4$\pm$$  7.1$ &  34.5$\pm$$ 2.9$ & 1.5$\pm$0.1 & 0.14$\pm$0.05 & 100 &  44 & \citet{wittenmyer2016b}\\
			&  &  &   &   &  &  1.5$\pm$0.2 &  851.7$\pm$$ 31.8$ &  26.7$\pm$$ 4.2$ & 2.0$\pm$0.1 & 0.00$\pm$0.07 & 100 &  29 & \citet{wittenmyer2016b}\\
			31592 & 1.4$\pm$0.2 & 3.2$\pm$0.1 &  11.6$\pm$$ 0.3$ &  0.21$\pm$0.10 & RGB &  1.9$\pm$0.1 &  744.1$\pm$$  3.6$ &  34.7$\pm$$ 1.3$ & 1.8$\pm$0.1 & 0.18$\pm$0.05 & 100 &  41 & \citet{wittenmyer2011b}\\
			&  &  &   &   &  &  0.8$\pm$0.1 & 1027.4$\pm$$ 15.8$ &  13.0$\pm$$ 1.3$ & 2.2$\pm$0.1 & 0.16$\pm$0.09 & 100 &   8 & \citet{luque2019}\\
			34693 & 2.5$\pm$0.2 & 2.0$\pm$0.1 & 229.5$\pm$$ 9.0$ &  0.14$\pm$0.10 & HB & 21.3$\pm$0.2 &  305.5$\pm$$  0.1$ & 350.2$\pm$$ 3.1$ & 1.2$\pm$0.1 & 0.03$\pm$0.01 & 100 &  99 & \citet{mitchell2013}\\
			36616 & 1.7$\pm$0.5 & 2.6$\pm$0.2 &  72.8$\pm$$ 3.0$ &  0.15$\pm$0.10 & HB &  6.4$\pm$0.2 &  299.4$\pm$$  0.2$ & 136.9$\pm$$ 2.9$ & 1.0$\pm$0.2 & 0.00$\pm$0.02 & 100 &  96 & \citet{ortiz2016}\\
			37826 & 2.1$\pm$0.2 & 2.9$\pm$0.1 &  39.2$\pm$$ 0.2$ &  0.07$\pm$0.10 & HB &  3.1$\pm$0.1 &  598.2$\pm$$  1.3$ &  46.1$\pm$$ 1.4$ & 1.8$\pm$0.1 & 0.06$\pm$0.03 & 100 &  76 & \citet{hatzes2006,reffert2006}\\
			39177 & 2.9$\pm$0.4 & 1.7$\pm$0.1 & 478.8$\pm$$97.9$ &  0.06$\pm$0.10 & HB & 20.8$\pm$1.0 &  853.6$\pm$$  4.4$ & 216.7$\pm$$ 9.8$ & 2.5$\pm$0.1 & 0.05$\pm$0.04 & 100 &  98 & \citet{tala2020}\\
			43803 & 1.1$\pm$0.1 & 2.7$\pm$0.1 &  30.2$\pm$$ 0.4$ & -0.11$\pm$0.10 & RGB &  3.9$\pm$0.3 &  416.3$\pm$$  0.3$ & 175.8$\pm$$13.2$ & 1.1$\pm$0.1 & 0.83$\pm$0.02 & 100 &  87 & \citet{wittenmyer2017b}\\
			49129 & 1.2$\pm$0.6 & 2.4$\pm$0.3 &  53.2$\pm$$ 1.1$ &  0.04$\pm$0.10 & HB &  2.9$\pm$0.3 & 1268.1$\pm$$ 35.5$ &  49.8$\pm$$ 5.2$ & 2.4$\pm$0.5 & 0.17$\pm$0.08 & 100 &  54 & \citet{wittenmyer2017a}\\
			56640 & 1.0$\pm$0.1 & 3.1$\pm$0.1 &   9.8$\pm$$ 1.2$ &  0.09$\pm$0.09 & RGB &  3.7$\pm$0.1 & 2596.8$\pm$$ 81.8$ &  53.8$\pm$$ 2.2$ & 3.7$\pm$0.2 & 0.15$\pm$0.05 & 100 &  41 & \citet{jones2020}\\
			60202 & 1.9$\pm$0.2 & 2.3$\pm$0.1 & 114.7$\pm$$ 4.2$ & -0.24$\pm$0.10 & HB & 15.2$\pm$0.4 &  323.6$\pm$$  0.8$ & 302.1$\pm$$ 8.0$ & 1.1$\pm$0.1 & 0.25$\pm$0.02 & 100 &  99 & \citet{liu2008}\\
			63242 & 1.1$\pm$0.1 & 2.6$\pm$0.1 &  31.7$\pm$$ 3.0$ & -0.31$\pm$0.09 & RGB &  6.4$\pm$0.2 &  125.9$\pm$$  0.1$ & 248.0$\pm$$ 7.8$ & 0.5$\pm$0.1 & 0.05$\pm$0.03 & 100 &  98 & \citet{jones2013}\\
			65891 & 1.9$\pm$0.1 & 2.9$\pm$0.1 &  36.2$\pm$$ 3.6$ &  0.16$\pm$0.10 & HB &  5.9$\pm$0.2 & 1108.5$\pm$$ 17.9$ &  74.5$\pm$$ 2.8$ & 2.6$\pm$0.1 & 0.04$\pm$0.03 & 100 &  86 & \citet{jones2015b}\\
			67537 & 1.9$\pm$0.1 & 2.9$\pm$0.1 &  34.8$\pm$$ 1.6$ &  0.15$\pm$0.08 & HB &  8.5$\pm$0.7 & 2390.6$\pm$$305.0$ & 107.8$\pm$$ 6.7$ & 4.3$\pm$0.4 & 0.64$\pm$0.07 &  98 &  79 & \citet{jones2017}\\
			67851 & 1.3$\pm$0.2 & 3.1$\pm$0.1 &  14.0$\pm$$ 0.6$ &  0.00$\pm$0.10 & RGB &  1.2$\pm$0.1 &   89.0$\pm$$  0.1$ &  44.9$\pm$$ 2.0$ & 0.4$\pm$0.1 & 0.13$\pm$0.04 & 100 &  63 & \citet{jones2015a}\\
			&  &  &   &   &  &  4.8$\pm$0.3 & 2123.1$\pm$$127.6$ &  62.6$\pm$$ 3.3$ & 3.5$\pm$0.2 & 0.03$\pm$0.05 & 100 &  64 & \citet{wittenmyer2015}\\
			74890 & 1.4$\pm$0.2 & 3.2$\pm$0.1 &  12.0$\pm$$ 1.0$ &  0.20$\pm$0.13 & RGB &  2.1$\pm$0.1 &  827.1$\pm$$ 14.9$ &  38.0$\pm$$ 1.9$ & 1.9$\pm$0.1 & 0.13$\pm$0.05 & 100 &  46 & \citet{jones2016}\\
			75092 & 1.2$\pm$0.1 & 3.3$\pm$0.1 &   8.0$\pm$$ 0.8$ &  0.09$\pm$0.11 & RGB &  1.5$\pm$0.1 & 1048.9$\pm$$ 38.0$ &  29.5$\pm$$ 2.4$ & 2.1$\pm$0.1 & 0.35$\pm$0.08 & 100 &  26 & \citet{jones2020}\\
			75458 & 1.3$\pm$0.2 & 2.4$\pm$0.1 &  57.2$\pm$$ 0.8$ &  0.11$\pm$0.10 & HB & 10.0$\pm$0.1 &  511.0$\pm$$  0.0$ & 307.6$\pm$$ 2.0$ & 1.4$\pm$0.1 & 0.72$\pm$0.00 & 100 &  98 & \citet{frink2002}\\
			79540 & 1.2$\pm$0.1 & 1.7$\pm$0.1 & 179.6$\pm$$11.1$ & -0.01$\pm$0.10 & RGB &  4.3$\pm$0.2 &  578.2$\pm$$  1.9$ &  91.5$\pm$$ 4.2$ & 1.4$\pm$0.1 & 0.05$\pm$0.04 & 100 &  87 & \citet{tala2020}\\
			84056 & 1.3$\pm$0.1 & 3.2$\pm$0.1 &  10.1$\pm$$ 0.7$ &  0.08$\pm$0.07 & RGB &  1.8$\pm$0.2 &  877.0$\pm$$ 19.9$ &  33.6$\pm$$ 2.9$ & 1.9$\pm$0.1 & 0.09$\pm$0.07 & 100 &  36 & \citet{wittenmyer2016a}\\
			88048 & 2.7$\pm$0.1 & 2.6$\pm$0.1 & 107.0$\pm$$ 2.3$ &  0.06$\pm$0.10 & HB & 22.4$\pm$0.1 &  530.0$\pm$$  0.1$ & 288.5$\pm$$ 1.1$ & 1.8$\pm$0.1 & 0.13$\pm$0.00 & 100 &  99 & \citet{quirrenbach2011}\\
			&  &  &  &   &  & 24.4$\pm$0.4 & 3168.1$\pm$$ 28.7$ & 174.4$\pm$$ 2.2$ & 5.9$\pm$0.1 & 0.19$\pm$0.01 & 100 &  84 & \citet{quirrenbach2011}\\
			90988 & 1.4$\pm$0.2 & 3.3$\pm$0.1 &   8.6$\pm$$ 1.1$ &  0.24$\pm$0.14 & RGB &  2.2$\pm$0.1 &  455.3$\pm$$  3.2$ &  49.3$\pm$$ 2.7$ & 1.3$\pm$0.1 & 0.17$\pm$0.05 & 100 &  63 & \citet{jones2020}\\
			95124 & 1.4$\pm$0.1 & 3.3$\pm$0.1 &  10.6$\pm$$ 1.0$ &  0.20$\pm$0.08 & RGB &  2.3$\pm$0.1 &  566.8$\pm$$  5.3$ &  44.7$\pm$$ 2.4$ & 1.5$\pm$0.1 & 0.08$\pm$0.06 & 100 &  62 & \citet{jones2016}\\
			105854 & 1.9$\pm$0.1 & 2.8$\pm$0.1 &  36.9$\pm$$ 1.6$ &  0.31$\pm$0.18 & HB &  7.5$\pm$0.3 &  184.9$\pm$$  0.4$ & 174.3$\pm$$ 6.0$ & 0.8$\pm$0.1 & 0.04$\pm$0.03 & 100 &  98 & \citet{jones2014}\\
			107773 & 1.7$\pm$0.4 & 2.6$\pm$0.2 &  58.8$\pm$$ 4.4$ &  0.03$\pm$0.10 & HB &  1.5$\pm$0.1 &  144.0$\pm$$  0.4$ &  42.1$\pm$$ 2.7$ & 0.6$\pm$0.1 & 0.11$\pm$0.05 & 100 &  68 & \citet{jones2015b}\\
			114855 & 1.1$\pm$0.1 & 2.4$\pm$0.1 &  50.7$\pm$$ 1.0$ & -0.03$\pm$0.10 & HB &  2.7$\pm$0.1 &  181.4$\pm$$  0.1$ &  91.6$\pm$$ 2.3$ & 0.6$\pm$0.1 & 0.03$\pm$0.02 & 100 &  87 & \citet{mitchell2013}\\
			114933 & 1.3$\pm$0.1 & 3.3$\pm$0.1 &   8.9$\pm$$ 1.1$ &  0.14$\pm$0.08 & RGB &  2.1$\pm$0.2 & 1502.5$\pm$$ 29.2$ &  31.4$\pm$$ 4.0$ & 2.8$\pm$0.1 & 0.26$\pm$0.09 & 100 &  32 & \citet{jones2020}\\
			116630 & 1.2$\pm$0.2 & 3.4$\pm$0.1 &   6.8$\pm$$ 0.5$ &  0.16$\pm$0.14 & RGB &  1.8$\pm$0.1 &  876.8$\pm$$ 14.3$ &  33.1$\pm$$ 2.4$ & 1.9$\pm$0.1 & 0.09$\pm$0.05 & 100 &  36 & \citet{wittenmyer2017a}\\
			\hline  \\ [-1.7ex]
			11791 & 1.9$\pm$0.2 & 2.8$\pm$0.1 &  45.3$\pm$$ 3.5$ &  0.01$\pm$0.08 & HB &  3.0 &  691.9$\pm$$  3.6$ &  38.3$\pm$$ 2.0$ & 2.1 & 0.12$\pm$0.05 &  &  & \citet{sato2012}\\
			31674 & 1.4$\pm$0.2 & 2.9$\pm$0.1 &  28.2$\pm$$ 0.4$ & -0.07$\pm$0.10 & RGB &  1.8$\pm$0.2 &  363.3$\pm$$  2.5$ &  33.6$\pm$$ 3.2$ & 1.2$\pm$0.1 & 0.09$\pm$0.07 &  &  & \citet{sato2016}\\
			&  &  &   &  &  &  1.9$\pm$0.2 &  684.7$\pm$$  5.0$ &  30.1$\pm$$ 2.0$ & 1.9$\pm$0.1 & 0.28$\pm$0.08 &  &  & \citet{sato2016}\\
			40526 & 1.5$\pm$0.2 & 1.2$\pm$0.1 & 632.5$\pm$$17.2$ & -0.19$\pm$0.10 & RGB &  7.8$\pm$0.8 &  605.2$\pm$$  4.0$ & 133.0$\pm$$ 8.8$ & 1.7$\pm$0.1 & 0.08$\pm$0.02 &  &  & \citet{lee2014b}\\
			53261 & 2.2$\pm$0.4 & 1.5$\pm$0.1 & 569.6$\pm$$56.3$ & -0.21$\pm$0.10 & HB & 12.1$\pm$0.6 &  785.7$\pm$$  4.8$ & 158.7$\pm$$ 8.6$ & 2.2$\pm$0.1 & 0.11$\pm$0.06 &  &  & \citet{tala2020}\\
			107089 & 1.4$\pm$0.1 & 3.2$\pm$0.1 &  13.3$\pm$$ 0.3$ &  0.08$\pm$0.10 & RGB &  2.1$\pm$0.1 &  414.8$\pm$$  3.1$ &  38.7$\pm$$ 1.1$ & 1.3$\pm$0.1 & 0.09$\pm$0.04 &  &  & \citet{ramm2009,ramm2016}\\
			108513 & 1.4$\pm$0.1 & 3.2$\pm$0.1 &  12.6$\pm$$ 0.2$ &  0.13$\pm$0.10 & RGB &  1.4$\pm$0.1 &  352.7$\pm$$  1.7$ &  34.7$\pm$$ 2.2$ & 1.0$\pm$0.1 & 0.07$\pm$0.06 &  &  & \citet{yilmaz2017}\\ [0.3ex] 
			\hline  \\ [-1.7ex]
			9884 & 1.6$\pm$0.3 & 2.3$\pm$0.1 &  81.2$\pm$$ 1.0$ & -0.13$\pm$0.10 & RGB &  1.8$\pm$0.2 &  380.8$\pm$$  0.3$ &  41.1$\pm$$ 0.8$ & 1.2 & 0.25$\pm$0.03 &  &  & \citet{lee2011}\\
			41704 & 2.7$\pm$0.1 & 2.5$\pm$0.1 & 138.1$\pm$$ 2.1$ & -0.16$\pm$0.10 & HB &  4.1 & 1630.0$\pm$$ 35.0$ &  33.6$\pm$$ 2.1$ & 3.9 & 0.13$\pm$0.06 &  &  & \citet{sato2012}\\
			48455 & 1.5$\pm$0.1 & 2.5$\pm$0.1 &  51.5$\pm$$ 0.7$ &  0.29$\pm$0.10 & HB &  2.4$\pm$0.4 &  357.8$\pm$$  1.2$ &  52.0$\pm$$ 5.4$ & 1.1$\pm$0.1 & 0.09$\pm$0.06 &  &  & \citet{lee2014b}\\
			70791 & 1.0$\pm$0.1 & 2.4$\pm$0.1 &  60.8$\pm$$ 2.2$ & -0.70$\pm$0.10 & HB &  0.9$\pm$0.1 &   30.4$\pm$$  0.1$ &  59.9$\pm$$ 3.2$ & 0.2$\pm$0.1 & 0.04$\pm$0.04 &  &  & \citet{takarada2018}\\ [0.3ex] 
			\hline
	\end{tabular}}
	\tablefoot{The 37 systems in the first block will be considered for the analysis. The host stars of the second block are removed by our cuts ($\#\mathrm{RVs}\le10$ or -- in the cases of HIP~40526 and HIP~53261 -- $\log(g)\le1.5$). For the four stars in the third block, our data do not allow for detection with comparable confidence to the systems in the first block; they are hence excluded from the analysis (this includes the outer planet in the multiple system around HIP~19011). The stellar parameters are derived with the method of \citet{stock2018}. For the first block, we quote the orbital parameters as determined from our RV data with \texttt{RVLIN} \citep{wright2009}, and the completeness fraction $f_\mathrm{c}$ and recovery rate $f_\mathrm{R}$ used for the incompleteness corrections (see Sec.~\ref{sec_occ_global}). Uncertainties on the orbital parameters are determined via bootstrapping using the \texttt{BOOTTRAN} package \citep{wang2012}. For completeness, we also list the orbital parameters of the planets in the second and third block as given in their respective reference. For simplicity, asymmetric uncertainties were averaged into a single error estimate.}
\end{sidewaystable*}

\begin{figure*}
	\centering 
	\includegraphics[width=8cm]{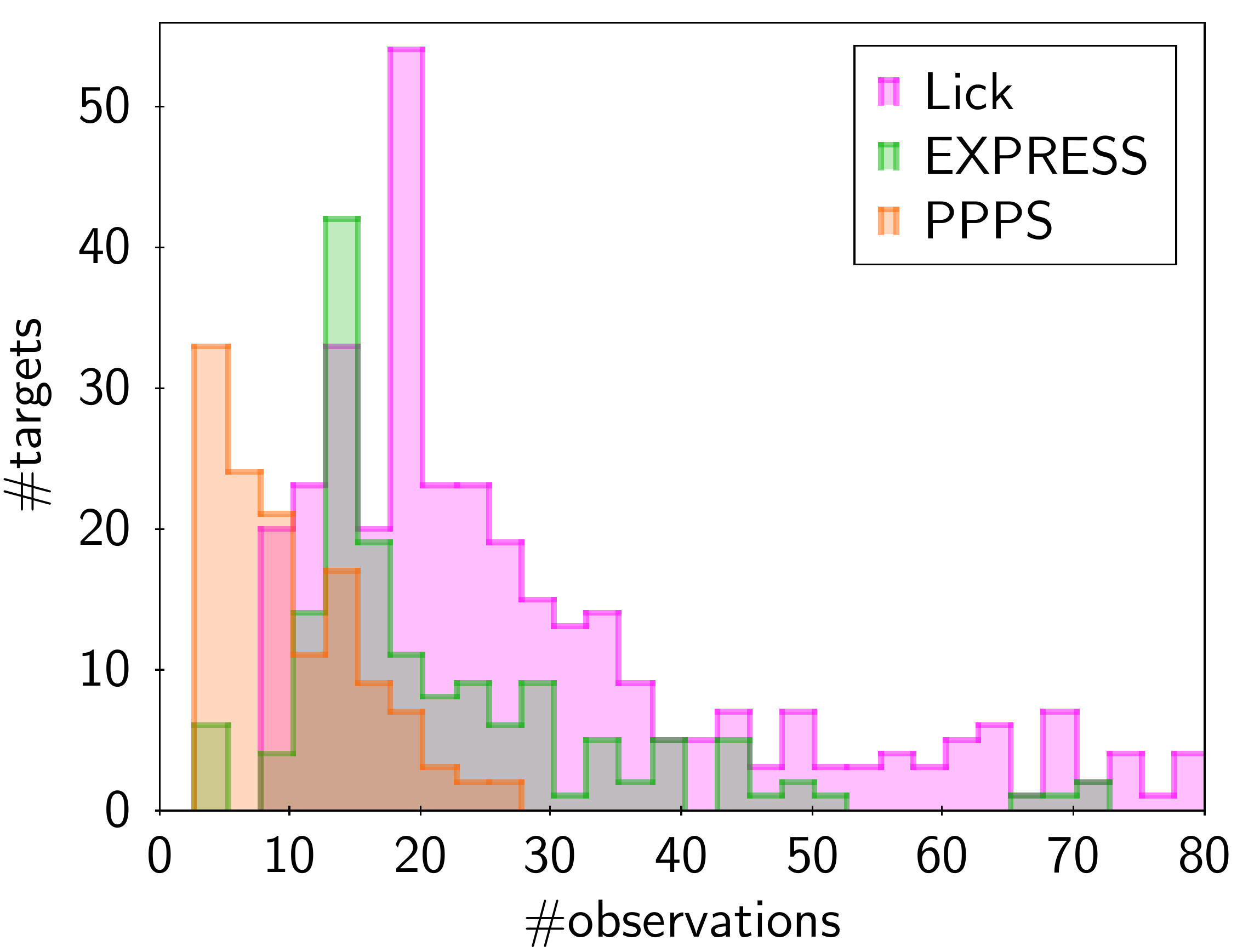}
	\includegraphics[width=8cm]{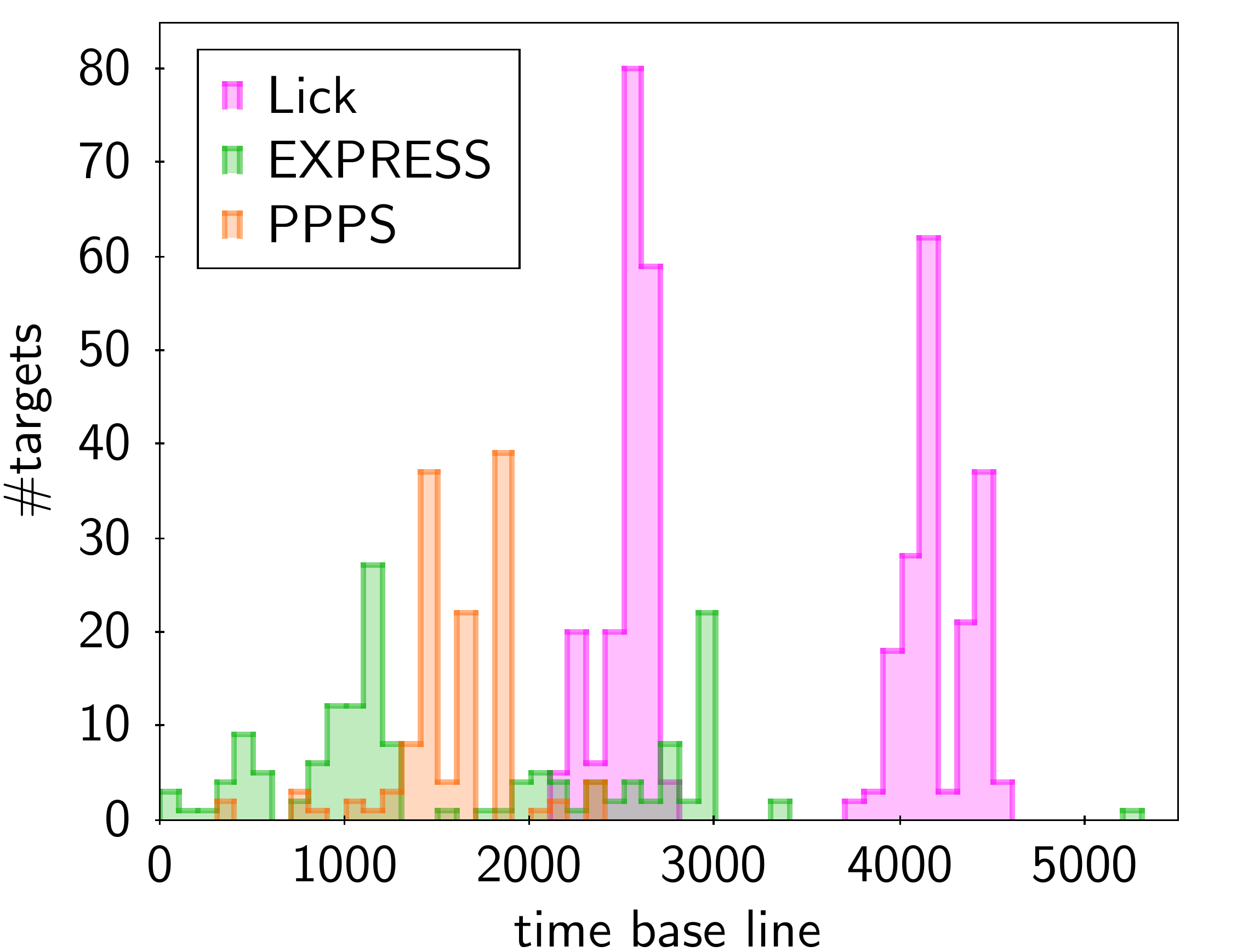}
	\caption{Histograms to illustrate the inhomogeneous observational histories among all stars of the three surveys. Lick, EXPRESS, and PPPS are shown in violet, green, and orange, respectively. \textit{Left}: Distribution of the number of radial velocity data points per star. \textit{Right}: Distribution of the time base line, i.e., the time span covered by RV observations per star.}
	\label{fig_hetero}
\end{figure*}

In total, 48 planets in 42 systems have been detected and published around all 650 stars from the target lists of Lick, EXPRESS, and PPPS. They are listed in Table~\ref{table_planets} including their detection reference. The stellar parameters given are the rederived ones (see Sec.~\ref{sec_sample}) and the minimum planet masses are based on the newly determined stellar masses. For confirmation of the planetary nature of the tabulated signals we refer to the listed discovery papers.\par 
Notable discoveries made by the three surveys include the first planet detected around a giant star, $\iota$~Dra~b \citep[HIP~75458~b; ][]{frink2002}, the most eccentric planet hosted by an evolved star, HD~76920~b \citep[HIP~43803~b; ][]{wittenmyer2017b,bergmann2021arXiv}, an intriguing multiple system containing one of the shortest-period planets ($P=89\,\mathrm{days}$, $a=0.4\,\mathrm{AU}$) detected around a giant star and a cool Jupiter at 3.5\,AU \citep[HIP~67851; ][]{jones2015a,jones2015b}, the S-type planet HD~59686~Ab in an eccentric, close-separation binary system \citep[HIP~36616; ][]{ortiz2016}, and several planets in mean motion resonances~(MMR): $\eta$~Cet~b and~c in a 2:1~MMR \citep[HIP~5364; ][]{trifonov2014}, HD~33844~b and~c  in a 3:5~MMR \citep[HIP~24275; ][]{wittenmyer2016b}, 7~CMa~b and~c in a 4:3~MMR \citep[HIP~31592; ][]{luque2019}, and $\nu$~Oph~b and~c in a 6:1~MMR \citep[HIP~88048; ][]{quirrenbach2019}.\par
We note that we include companions beyond the deuterium burning limit at 13\,$M_\mathrm{Jup}$ up to about 30\,$M_\mathrm{Jup}$ in the category ``planet'', as the high-order MMR of $\nu$~Oph~b and~c with their minimum masses above 20\,$M_\mathrm{Jup}$ strongly indicates that they formed in the circumstellar disk of $\nu$~Oph \citep{quirrenbach2019}. Within the combined sample, 79 stars have stellar binary companions ($M>100\,M_\mathrm{Jup}$). They are not be excluded from our analysis as HD~59686~Ab demonstrates that they can harbor detectable planetary systems.\par
Table~\ref{table_planets} also contains planets published by other RV giant star surveys. For the sake of our occurrence rate analysis, we only consider these planets if their observational history within our combined survey allows for a detection with comparable confidence to the planets published within the framework of the Lick, EXPRESS, and PPPS programs. We therefore divide the entries in Table~\ref{table_planets} into three blocks: the first block encompasses planets with clear evidence in our RV data and which are the basis for the occurrence rate analysis; 6~planets are hosted by stars that are removed from the sample by the homogenizing cuts described in Sec.~\ref{sec_sample} -- these are assigned to the second block; the third block finally lists those planets without sufficient evidence in our data. While we do find hints of the signals from the third block, the observations are not enough to fully constrain the orbits and/or the false alarm probability is larger than 1\%. Thus, they do not differ from other potential planets that might still be hidden in the data and we treat them as such.\par
Some candidates from the PPPS whose orbits are still unconstrained and follow-up observations are still needed to confirm them are presented in \citet{wittenmyer2020}. As the planets from block three and the candidates are only excluded due to insufficient observations, they should largely be automatically absorbed by the completeness correction for the occurrence rates.\par
In summary, we work with a sample of 37 planets in 32 systems (13.5\% rate of observed multiples) in the following sections, spanning a range from 0.8\,$M_\mathrm{Jup}$ to 24.4\,$M_\mathrm{Jup}$ in minimum planet mass and from 89\,days to 3168\,days in period.

\section{Detection probabilities} \label{sec_detprobs}
The observational coverage and the noise properties of each star differ considerably across the combined survey. Figure~\ref{fig_hetero} illustrates how the number of data points and length of observation vary for the individual stars within the three surveys. To properly account for the imperfect observations, we use injection-recovery tests to determine the detection capabilities. Our method is a modification of the periodogram analysis used by, for example, \citet{endl2002}, \citet{zechmeister2009}, and \citet{mayor2011}, but accounting for quasi-periodic stellar noise, which several of our giant stars exhibit and which is distinct from the well-known p-mode oscillations (which have timescales of $\sim$hours-days, are not resolved by our observing cadence, and therefore show up as noise). The origin of these longer-period oscillations is not yet well understood (see also Sec.~\ref{sec_occ_period}).
\begin{figure*}
	\centering
	\includegraphics[width=8cm]{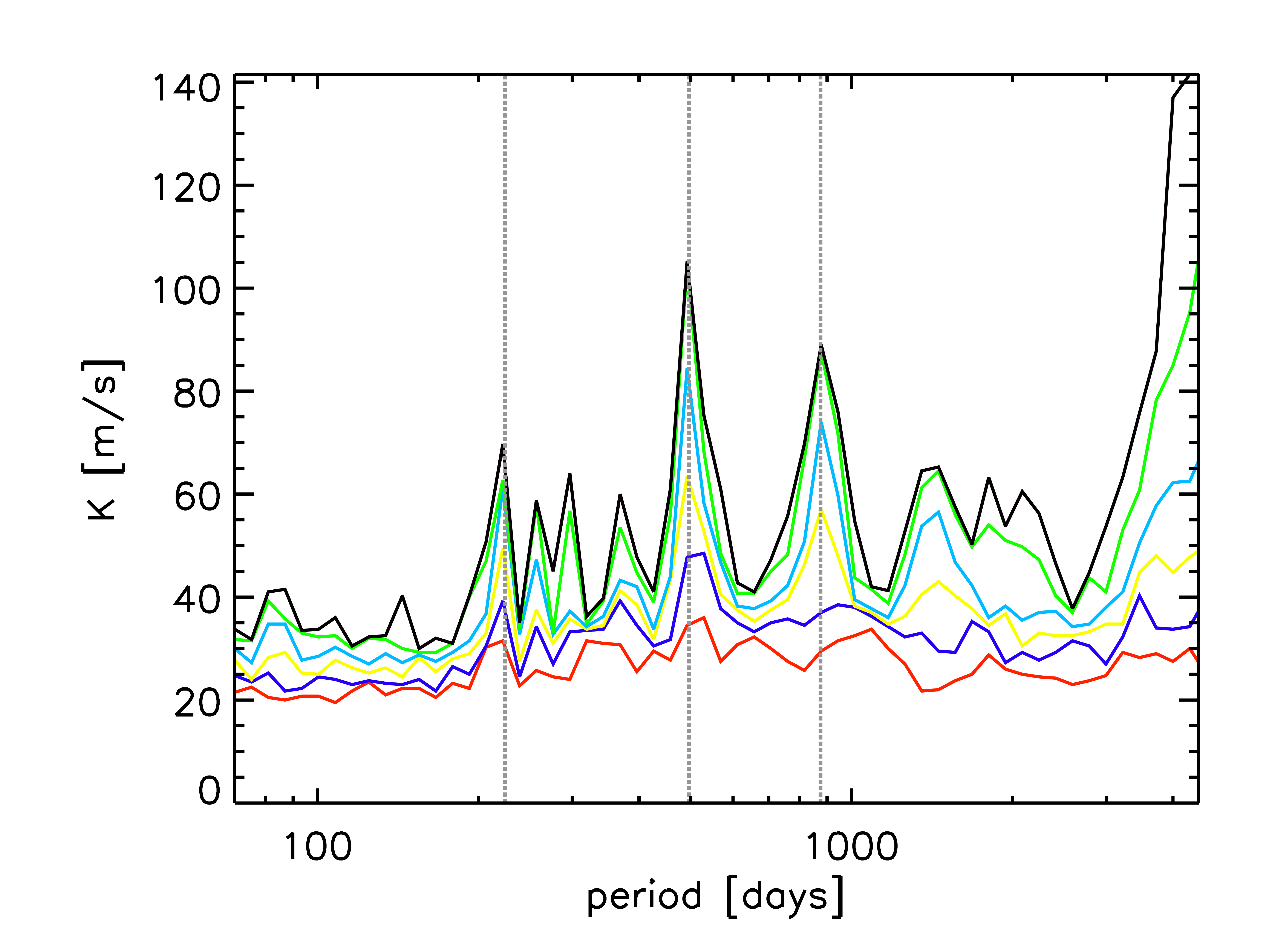}
	\includegraphics[width=8cm]{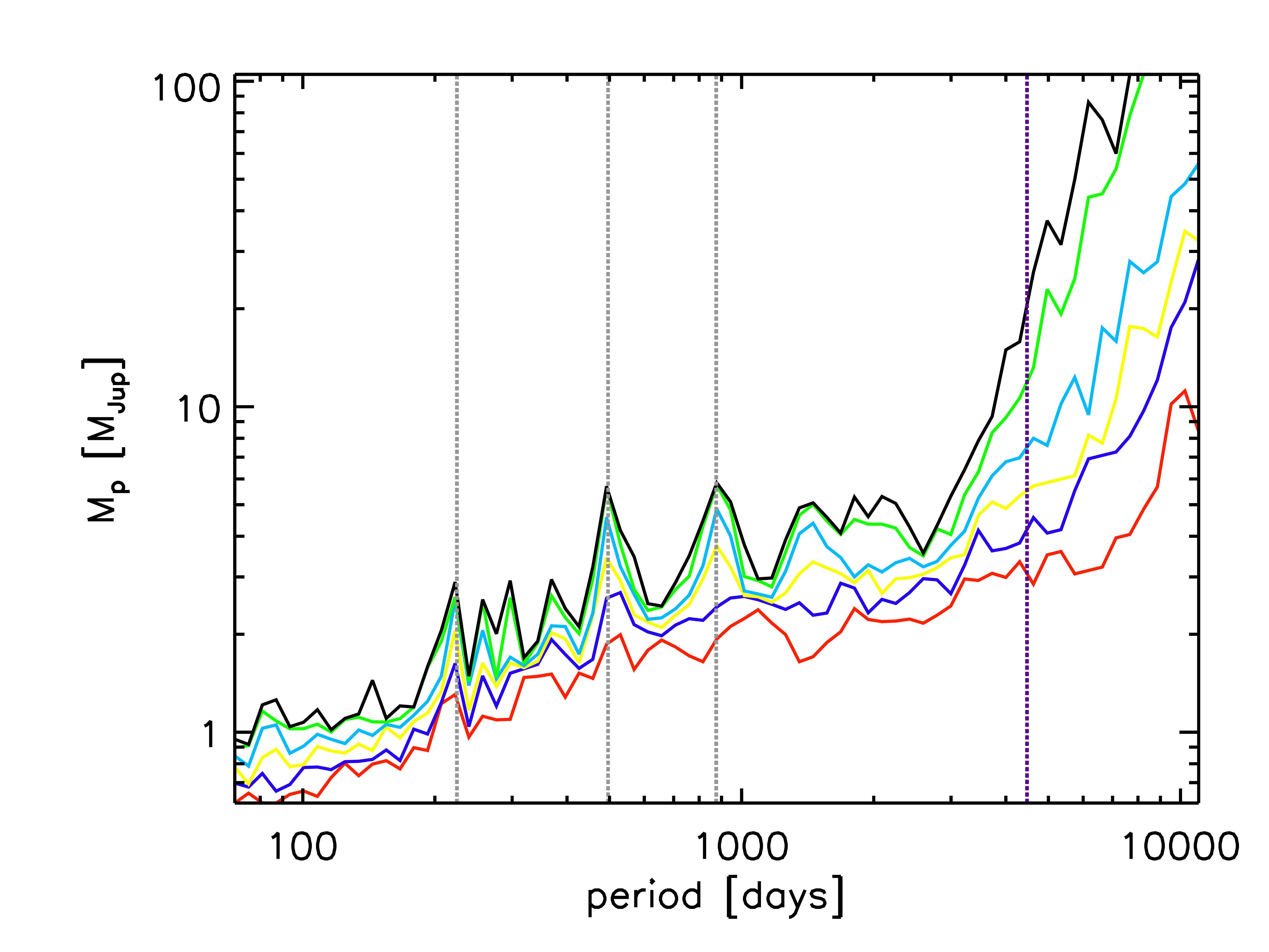}
	\caption{Detection probability map for a target star affected by quasi-periodic oscillations; it exhibits three peaks (at 224d, 496d, and 875d) with FAP<1\% in the GLS periodogram; \textit{left}: in period-RV semi-amplitude space, and \textit{right:} in period-planet mass space. The black, green, light blue, yellow, blue, and red solid lines correspond to detection probabilities of 100\%, 90\%, 70\%, 50\%, 30\%, and 10\%, respectively. The detection probability level of, e.g., 10\% gives the semi-amplitude (or planet mass) at which 10\% of the tested phases can be recovered at the given period. The vertical black dotted lines show the positions of the three significant periodicities, while the vertical dashed blue line in the right plot delineates the time base line of observations for this star (the left plot is restricted to the time base line for better visibility).}
	\label{fig_example}
\end{figure*}

\begin{figure*}
	\centering
	\includegraphics[width=8cm]{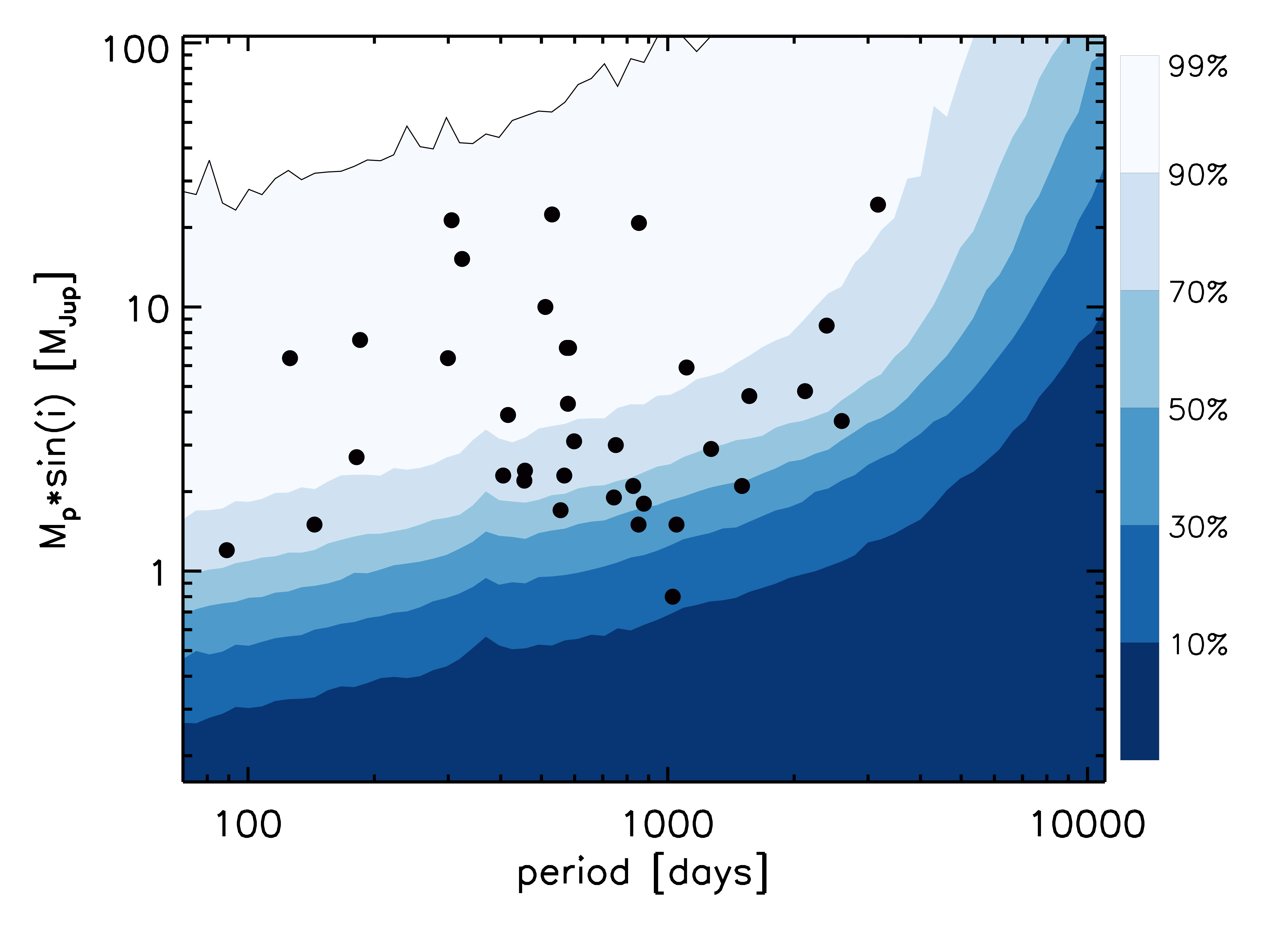}
	\includegraphics[width=8cm]{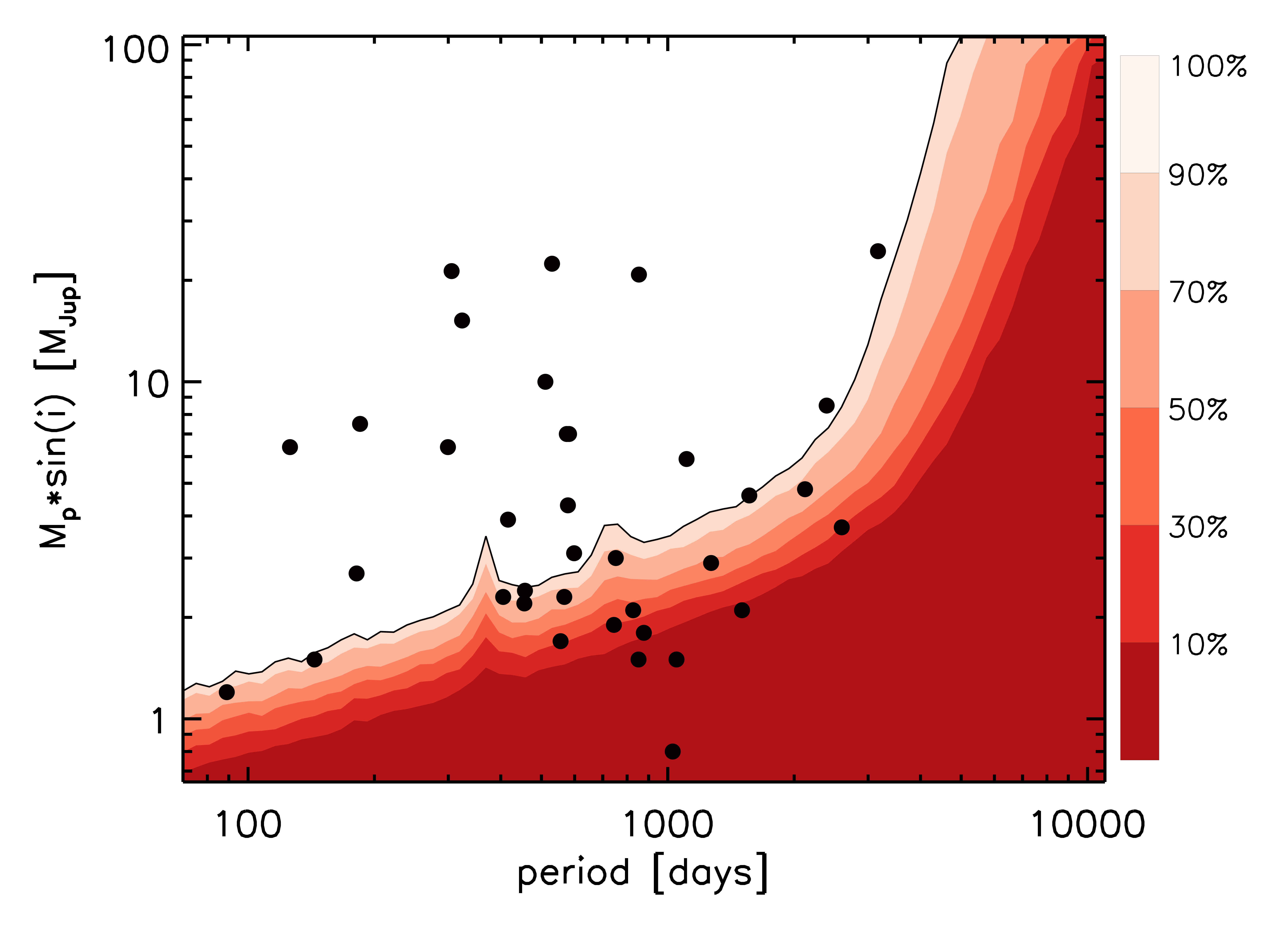}
	\caption{Two different ways to visualize detectability within the survey. Detected planets (see Table~\ref{table_planets}) are plotted in black. 
		\textit{Left}: Global detection efficiency of the combined survey in period-planet mass space. Planets with $M_p\sin(i)$ above the, e.g., 50\% limit can be detected around 50\% of the stars in the sample at the given period.
		\textit{Right}: Median detection probabilities of the combined survey in period-planet mass space (the 50\% median probability curve is computed by determining the median of all individual stars' 50\% level at each period). Planets with $M_p\sin(i)$ above the 50\% limit have a median detection probability of 50\% across the sample at the given period.}
	\label{fig_efficiency}
\end{figure*}

\subsection{Injection-recovery tests}
We start by removing the Keplerians of all 37~confidently detected planets (first block in Table~\ref{table_planets}) and 79~binary companions from the RV time series of the respective host stars using the orbit-fitting routine \texttt{RVLIN}, which is based on a partial linearization of Kepler's equations \citep{wright2009,rvlin}. The resulting data can then be considered as pure noise, which can be caused by stellar ``jitter'', measurement noise, yet undetected planets, or a combination thereof. Also the 4~planets from the third block of Table~\ref{table_planets} are treated as part of this noise. In particular, we note that evolved stars often exhibit quasi-periodic variations (which can have a false-alarm-probability (FAP) of less than 1\%) that we also consider as part of the noise and which require us to make some adaptations to the commonly used periodogram analysis.\par
Before adding synthetic planets with different periods and masses to the resulting residuals, we determine the power threshold in the GLS periodogram \citep{gls_paper} corresponding to a FAP level of 1\% using bootstrap randomization \citep{kuerster1997}: For each star, we construct 10\,000 bootstrapped data sets by shuffling the measured RV data points while keeping the observing times fixed. We then compute the GLS for each data set and record the maximum power. The 1\%~FAP level is given by the 99th percentile of the resulting distribution of maximum powers.\par
In the next step, we mimic planetary signals by adding sine waves with different periods, RV semi-amplitudes, and orbital phases to the residuals of each star. We limit our detectability analysis to circular orbits as the impact of the eccentricity on the resulting detection limits is relatively small for eccentricities up to about 0.5 \citep[see][]{endl2002, cumming2008,cummingdragomir2010}.
For each star, we probe 70~trial periods ranging from 70~days to 11\,000~days and 50 equidistant phases, increasing the signal's semi-amplitude until the planet becomes detectable or an amplitude corresponding to a stellar mass companion ($100~M_\mathrm{Jup}$) for the particular star is reached \citep[][]{wittenmyer2011a}. In order to deem a signal successfully recovered, we require the following criteria concerning peak height and position, as well as phase and amplitude recovery to be met.\par
We evaluate the peaks in the vicinity -- defined as five times the average peak width in the periodogram, which is given by the inverse of the time base line \citep{vanderplas2018} -- of the trial period and require that the power of the peak closest to the trial period rises above the previously determined 1\% FAP level, and that it is the highest peak in the region. This ensures that the synthetic peak rises above a potential peak in the vicinity corresponding to a (significant) stellar periodicity, whereas limiting the evaluated period range avoids spurious detections while the simulated signal is still small.\par
As mentioned already, evolved stars can exhibit stellar periodic signals with a FAP below 1\%. If the injected signal's period comes close to one of these noise periods, the first criterion may accept the signal as detected already at very small amplitudes because the stellar signal fulfills the requirements. Thus, it is essential to also test if the phase can be correctly retrieved. As the injected and recovered periods may differ slightly, the phase parameters cannot be compared directly. Instead, we determine the deviation of the recovered phase by finding the maximum separation between corresponding extrema of the two curves during the RV time series, while making sure that at least a time range of three periods is probed (which for long periods can extrapolate beyond the time base line). We then require that the deviation is smaller than 15\% of the trial period.\par
If the signal is recovered in all 50 phases, we define the reached semi-amplitude as the 100\% detection limit at the considered period. However, signals might still be detectable below this limit in a fraction of the tested phases due to fortunate sampling of the RV curve. We therefore assign detection probability levels of 10\%, 20\%, ... to the semi-amplitudes where the signal first became detectable in 10\%, 20\%, ... of phases. Yet again, it can happen that for low probabilities a stellar periodicity is recovered at small injected amplitudes when both period \emph{and phase} are sufficiently similar (as we are probing a smaller range of phases). Hence, we additionally require that the recovered amplitude is within 30\% of the simulated amplitude.\par 
This procedure provides us with detection maps for every star in semi-amplitude space, which can be translated into planet mass space using the homogeneously determined stellar masses -- an example is shown in Fig.~\ref{fig_example}. This star exhibits three peaks with FAP<1\% in its periodogram of unclear, most likely stellar origin and which we therefore treat as noise. As can be seen in the figure, the detection probability curves rise around these periods indicating that a stronger planetary signal is necessary to clearly distinguish it from stellar noise. We note that this star constitutes an extreme example and is not representative of the entire sample, but serves as a good showcase for how our procedure deals with (quasi-)periodic noise.
\subsection{Survey completeness}
The individual detection maps can be combined to illustrate the overall detection capabilities of the survey. The global survey efficiency (shown on the left side of Fig.~\ref{fig_efficiency}) takes into account the upper limits of each star by counting for each period which fraction of the sample stars confidently excludes planets above a certain mass, that is planets above the 50\% efficiency curve can be detected (or excluded) with a probability of 100\% around 50\% of the sample. The right side of Fig.~\ref{fig_efficiency} depicts the median detection probability curves which are computed by averaging each individual probability level over the entire sample, that is planets above the 50\% mean probability curve have an overall probability of being detected of 50\% across the sample (or are excluded with 50\% confidence). The different global efficiency curves spread across a larger range in planetary masses than the mean detectability curves because they depend more strongly on individual stars and their observational history. Both versions of detectability curves can give an idea about how many planets might have been missed by our observations in different domains of the planet mass-period space. For instance, many of the detected planets with masses smaller than $\sim$\,$3M_\mathrm{Jup}$ lie in regions of low probability indicating that more planets of this kind are to be expected and have remained elusive due to insufficient observations. Thus, their occurrence rate is considerably higher than the rate of detection alone suggests. In Sec.~\ref{sec_occ_period}, we look at the planet occurrence across different regions of planet mass and period space in more detail.

\section{Global giant planet occurrence around evolved stars} \label{sec_occ_global}
\subsection{Global occurrence rate of the combined sample}
If all stars in the sample were equally well observed and had the same noise properties, and the detected planets were all located above the upper detection limits, we could calculate the planet occurrence rate in the covered period and planet mass range simply by dividing the number of detections by the number of surveyed stars. As this is not the case, it is crucial to correct for incompleteness by estimating how many planets might have been missed using the detection probability maps from the injection-recovery tests. We follow the method for correction described by \citet{wittenmyer2011a}:\par
We start by computing the completeness fraction $f_\mathrm{c}\left(P_i,M_{\mathrm{P},i}\right)$ which represents the mean detection probability across the sample for a planet $i$ with given period $P_i$ and minimum mass $M_{\mathrm{P},i}$:
\begin{equation}
	f_\mathrm{c}(P_i,M_{\mathrm{P},i})=\frac{1}{n_\mathrm{stars}}\cdot\sum_{j=1}^{n_\mathrm{stars}} f_{\mathrm{R},j}\left( P_i,M_{\mathrm{P},i}\right)
	\label{eq_fc}
\end{equation}
where $n_\mathrm{stars}$ denotes the number of sample stars (after cuts) and $f_{\mathrm{R},j}\left(P_i,M_{\mathrm{P},i}\right)$ is the recovery fraction of planet $i$ around the star $j$, which is the probability for detecting planet $i$ if it was hosted by star $j$ given the quality and quantity of our observations of star $j$. This fraction is determined from the individual detection maps by interpolating between the enclosing detection probability curves (which were recorded in steps of 10\%) and the trial periods to obtain the probability corresponding to the values $P_i$ and $M_{\mathrm{P},i}$. Both $f_\mathrm{c}(P_i,M_{\mathrm{P},i})$ and $f_{\mathrm{R}}(P_i,M_{\mathrm{P},i})$ for all 37 planets can be found in Table~\ref{table_planets}.\par
We then combine the completeness fraction for each planet with the recovery fraction around its own host star, and sum over all host stars\footnote{For stars hosting multiple planets, we use the recovery fraction of the most easily detectable planet.} to calculate the number of planets missed due to insufficient observations:
\begin{equation}
	n_\mathrm{missed}=\sum_{i=1}^{n_\mathrm{hosts}}\frac{1}{f_{\mathrm{R},i}\left(P_i,M_{\mathrm{P},i}\right) \cdot f_\mathrm{c}\left(P_i,M_{\mathrm{P},i}\right) }-n_\mathrm{hosts}
\end{equation}
where $n_\mathrm{hosts}$ is the number of stars hosting at least one planet. We find that about 19.7 planets have been missed across the combined sample.\par
Of the 482 stars in our sample (after cuts) 32 were found to host at least one giant planet ($M_\mathrm{P}\ge 0.8$\,$M_\mathrm{Jup}$). Using binomial statistics, we get a first estimate of the occurrence rate of planetary systems in our sample (including its 68\% confidence interval from the beta distribution \citep{cameron2011}) as $f_\mathrm{binom}=n_\mathrm{hosts}/n_\mathrm{stars}=32/482=6.6\%^{+1.3\%}_{-1.0\%}$. 
We then use the number of missed detections to boost this estimate and its uncertainties according to
\begin{equation}
	f_\mathrm{occ}=f_\mathrm{binom}\cdot\frac{n_\mathrm{missed}+n_\mathrm{hosts}}{n_\mathrm{hosts}}
	\label{eq_focc} \text{ .}
\end{equation}
This finally gives us the corrected occurrence rate\footnote{Note that the corrected occurrence rate may also be computed directly: $f_\mathrm{occ}=\frac{1}{n_\mathrm{stars}}\cdot\sum_{i=1}^{n_\mathrm{hosts}}\frac{1}{f_{\mathrm{R},i}\left(P_i,M_{\mathrm{P},i}\right) \cdot f_\mathrm{c}\left(P_i,M_{\mathrm{P},i}\right) }$} of planetary systems hosting at least one giant planet within our sample:
\begin{equation*}
	f_\mathrm{occ} = 10.7\%^{+2.2\%}_{-1.6\%}
\end{equation*}
This agrees with the value of $7.8\%^{+9.1\%}_{-3.3\%}$ found for the PPPS sample alone \citep{wittenmyer2020}.\par 
\citet{cumming2008} estimate an occurrence rate of 6$-$9\% for planets with masses $M_\mathrm{P}=1-10$\,$M_\mathrm{Jup}$ and periods in the range from 2\,days to 10\,years. Compared to these observational results for solar-mass, main-sequence stars, we find a slightly larger occurrence rate, which can be explained by the fact that our sample encompasses more massive stars (see Sec.~\ref{sec_occ_mass}). Our global occurrence rate is at the lower edge of the rate of ~10-15\% predicted by theoretical studies (Kennedy \& Kenyon 2008), but we note that they look at gas giants in general and planets lighter than Jupiter are hardly detectable in our sample -- so we expect our occurrence rate to be smaller.\par
\citet{bowler2010} find the global planet occurrence of their subgiant sample to lie at $26\%^{+9\%}_{-8\%}$, which is significantly higher than our estimate. However, \citet{johnson2010} note that their sample of retired A stars is comparatively metal-rich which can cause a higher occurrence rate due to the planet-metallicity correlation. From Fig.~2 in \citet{johnson2010}, we can see that the average metallicity of the subgiant sample is around 0.15~dex, whereas our sample mean lies at $-0.05$~dex. We derive the exact shape of the planet-metallicity correlation for our sample in Sec.~\ref{sec_occ_mass} (taking also the effect of the stellar mass into account) and show its functional form in Fig.~\ref{fig_bayes}. For 0.15~dex and $-0.05$~dex and the mean stellar mass of our sample, our model places the expected occurrence rates at 20.2\% and 10.7\%, respectively -- hence our occurrence rate and the one derived by \cite{bowler2010} are fully consistent when factoring in the different stellar metallicities.\par
Upon the publication of four newly discovered planets (see also Table~\ref{table_planets}), \citet{jones2020} analyzed the fraction of stars hosting at least one giant planet in the common sample of EXPRESS and PPPS targets (37 stars, but four of them excluded from the analysis as they are compact binary systems; 11 planetary systems), finding a global occurrence rate of $33.3\%^{+9.0\%}_{-7.1\%}$, which again is significantly higher than what we find for our combined sample of three surveys.\footnote{Correcting this value for completeness as described above, we even find $36.8\%\pm0.1\%$. Only 1.16 planets are predicted to have been missed by the observations reflecting the fact that this common sample has been more thoroughly observed than the entire sample, on average.} In order to check whether the discrepancy can be explained by differing metallicity and/or stellar mass distributions between the common and the combined sample, we again anticipate our results from Sec.~\ref{sec_occ_mass}: We use the individual masses and metallicities to compute the expected occurrence rate for each of the 33 stars according to Eq.~\ref{R15}, then average these values to obtain the expected global occurrence rate, which yields 14.3\%. So indeed, the parameter distributions are in favor of a higher occurrence rate, but the obtained value is still much smaller than the observed one. In principle, such a result could occur randomly due to the limited sample size. We can use the Poisson binomial distribution to assess the probability of obtaining this result. Given the set of individual success probabilities from Eq.~\ref{R15}, its peak indicates that the most probable outcome is that 4 planets are distributed among the 33 stars with a probability of 20\% -- but with a probability of more than 50\%, the distribution function predicts a result of 5 or more planets in the subsample. However, the probability of 11 or more successes in 33 trials is only 0.4\%. In other words, considering the mean of the distribution function at 4.72 and its standard deviation of 1.98, the observed result for the common sample lies just outside the 3$\sigma$ confidence interval. We note that the stellar masses of the common sample are all $\leq$\,$1.42\,M_\sun$, and 14 out of the 33 stars and 8 out of the 11 planet hosts are found in the mass range of the bin with the highest occurrence rate depicted in Fig.~\ref{fig_bayes} -- and it is readily apparent that precisely this mass bin shows an excess of planet hosts with respect to the model prediction. Indeed, this excess is also visible in the Lick sample: Figure~3 of \citet{reffert2015} shows a high, narrow peak of $\sim$40\% in planet occurrence rate at metallicities around 0.2 and stellar masses just above $1.5\,M_\sun$, which appears to be distinct from the pronounced peak they find at $\sim$$2.0\,M_\sun$. While the underlying reason for the discrepancy remains elusive at the moment, it might hint at another process or effect that could be at work in this mass range, perhaps related to stellar evolution as all stars in the common sample are more likely to be RGB stars and seem to be at a fairly early stage in their evolution with radii smaller than $7.5\,R_\sun$ and $3.4\ge\log(g)>2.7$, so some of them could even be subgiants,.\par
\subsection{Global occurrence rate as function of evolutionary stage}
As mentioned before, stellar evolution can have an influence on orbiting bodies and may cause the planet occurrence rate to change. In our sample, 171 stars have a higher probability of still being on the first ascent of the red giant branch, while 311 stars are more likely to have already reached the horizontal branch. The planet hosts are divided equally: 16 of them are RGB stars (hosting in total 19 planets) and the other 16 are HB stars (hosting 18 planets). We derive the completeness corrected occurrence rate for both evolutionary stages as above and find:
\begin{equation*}
	f_\mathrm{occ,RGB} = 14.2\%^{+4.1\%}_{-2.7\%} \text{ \ \ and \ \ } f_\mathrm{occ,HB} = 6.6\%^{+2.0\%}_{-1.3\%}
\end{equation*}
For the RGB and HB stars, 8.3 and 4.4 planets have been missed, respectively.\par 
The occurrence rate of planetary systems on the RGB is more than twice as high as on the HB. In principle, this trend is in line with the expectation that the occurrence rate drops after a star has reached its maximum RGB radius as it could have engulfed potential close-in planets. Moreover, stellar mass loss can cause longer period orbits (that are too large to cause significant tidal interaction between star and companion) to expand leading to planets escaping detection. On the other hand, tides raised on the star by the accompanying body can cause orbits to shrink -- the outcome strongly depends on the strength of dissipation within the star, which is not very well constrained. However, the RGB and HB stars in our sample differ from each other in another aspect that might affect the occurrence rate more than the evolutionary stage: The stellar mass distribution is different in both subsamples, which is illustrated in the top right plot of Fig.~\ref{fig_cuts}. The RGB stars are almost exclusively confined to masses <2\,$M_\odot$, with a mean mass of 1.3\,$M_\odot$ (standard deviation 0.4\,$M_\odot$), whereas the HB stars have a more uniform mass distribution between 1\,$M_\odot$ and about 3.5\,$M_\odot$ with a small peak at $\sim$2.5\,$M_\odot$, the mean lies at 2.1\,$M_\odot$ (standard deviation 0.8\,$M_\odot$). These different mass distributions partly reflect the inherently different mass distributions of the observable population of RGB and HB stars in the Galaxy: stars with masses above the transition mass between degenerate and nondegenerate cores ascend the RGB significantly quicker than lower-mass stars as core contraction is not slowed by degeneracy pressure. At the same time, the HB lifetime peaks for stellar masses just above the transition mass while the RGB lifetime increases quickly toward lower masses \citep[see e.g., Fig.~6 of][]{girardi2013}. Thus, we are much more likely to observe an intermediate-mass star on the HB and a star below the transition mass on the RGB. The mass distributions in our sample however do not entirely follow those of the RGB and HB populations of the Galaxy, because the selection process involved limits on photometric stability and color -- one subset in particular was specifically selected to contain the highest-mass giants causing these to be over-represented in our sample compared to the Galactic population.\par
As we show in Sec.~\ref{sec_occ_mass}, the planet occurrence rate around evolved stars peaks at $\sim$1.7\,$M_\odot$, so the RGB stars are on average located closer to this maximum -- which could also be an explanation for the higher planet occurrence rate around RGB stars. Hence, we investigate the effects of both mass and evolutionary stage in shaping the occurrence rate further in Sec.~\ref{sec_occ_mass}.

\section{Planet occurrence as a function of period} \label{sec_occ_period}
\begin{figure}
	\resizebox{\hsize}{!}{\includegraphics{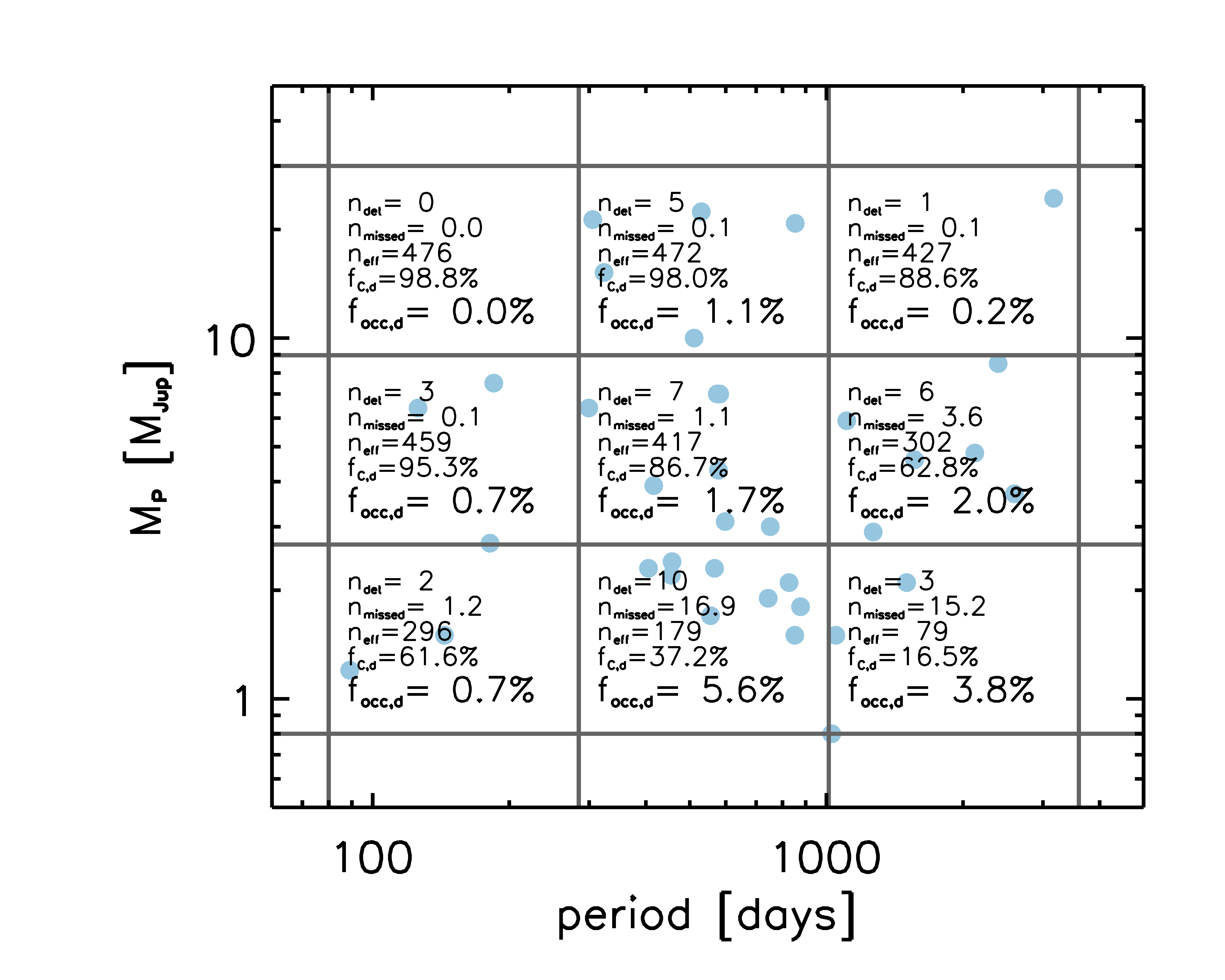}}
	\caption{Planet occurrence $f_\mathrm{occ,d}$ as a function of period and minimum planet mass. For each of the nine domains, we also give the number of detected planet $n_\mathrm{det}$, the estimated number of missed planets due to incompleteness $n_\mathrm{missed}$, the effective number of sample stars that are useful for constraining the occurrence rate in this domain $n_\mathrm{eff}$, and the mean completeness within the domain $f_\mathrm{C,d}$. The light blue dots show the positions of the detected planets.}
	\label{fig_mpplane}
\end{figure}
\subsection{Planet occurrence across the $P-M_p\sin(i)$ plane}
To investigate how the occurrence of planets varies with planetary mass and orbital period, we divide the region in the $M_\mathrm{P}\sin(i)-P$ diagram covered by detected planets into nine domains equal in $\log(M_\mathrm{P})$ and $\log(P)$ between 0.8 and 30\,$M_\mathrm{Jup}$ and 80 and 3600\,days, as shown in Fig.~\ref{fig_mpplane}. For each domain, we determine the mean completeness $f_\mathrm{C,d}$ by evaluating Eq.~\ref{eq_fc} for 2500 combinations of $\{P,M_\mathrm{P}\sin(i)\}$ across a $50\times 50$ log-uniform grid. We then use this to compute the number of missed planets in this domain: $n_\mathrm{missed}=n_\mathrm{det}\cdot \left(f_\mathrm{C,d}^{-1}-1 \right)$, where $n_\mathrm{det}$ is the number of detected planets in the domain. As we are interested in the fraction of stars with at least one planet in the given domain, we count each planet of multiple systems in its respective domain. Additionally, we compute $n_\mathrm{eff}=n_\mathrm{stars}\cdot f_\mathrm{C,d}$ which estimates the effective number of stars whose observations exclude planets that would be located in this domain. Finally, the occurrence rate $f_\mathrm{occ,d}$ is given by Eq.~\ref{eq_focc} with the substitution of $n_\mathrm{hosts}$ with $n_\mathrm{det}$. Figure~\ref{fig_mpplane} reports all four values for each of the nine domains. As can be seen, the most common planets with masses above $0.8\,M_\mathrm{Jup}$ around evolved stars have masses below 3\,$M_\mathrm{Jup}$ and are preferentially found at periods between several hundred days and $\sim$1000\,days. Furthermore, planets with long periods of a few thousand days are more abundant than periods below $\sim$300\,days in this mass regime. It seems that planets across the entire probed mass range are least common in the latter period regime, with planets above 10\,$M_\mathrm{Jup}$ being decidedly rare. In fact, we did not find a single one in this domain even though our completeness is highest. From the binomial distribution, we derive a 1$\sigma$ upper limit of $f_\mathrm{occ}\le 0.004$ for zero detections around 476 stars. \par 
\citet{cumming2008} find that the orbital period distribution shows an increase in the occurrence rate of gas giants of a factor of 5 beyond $P$$\sim$300\,days. And indeed, if we collapse Fig.~\ref{fig_mpplane} along the mass axis, we find for the three bins: $f_\mathrm{occ,1}=1.22\%^{+0.81\%}_{-0.34\%}$, $f_\mathrm{occ,2}=6.20\%^{+1.56\%}_{-1.05\%}$ and $f_\mathrm{occ,3}=3.75\%^{+1.56\%}_{-0.86\%}$, which corresponds to a factor of 5 increase between the first and the second. For periods above $\sim$1000\,days, the occurrence rate drops again by 40\%.
\begin{figure}
	\resizebox{\hsize}{!}{\includegraphics{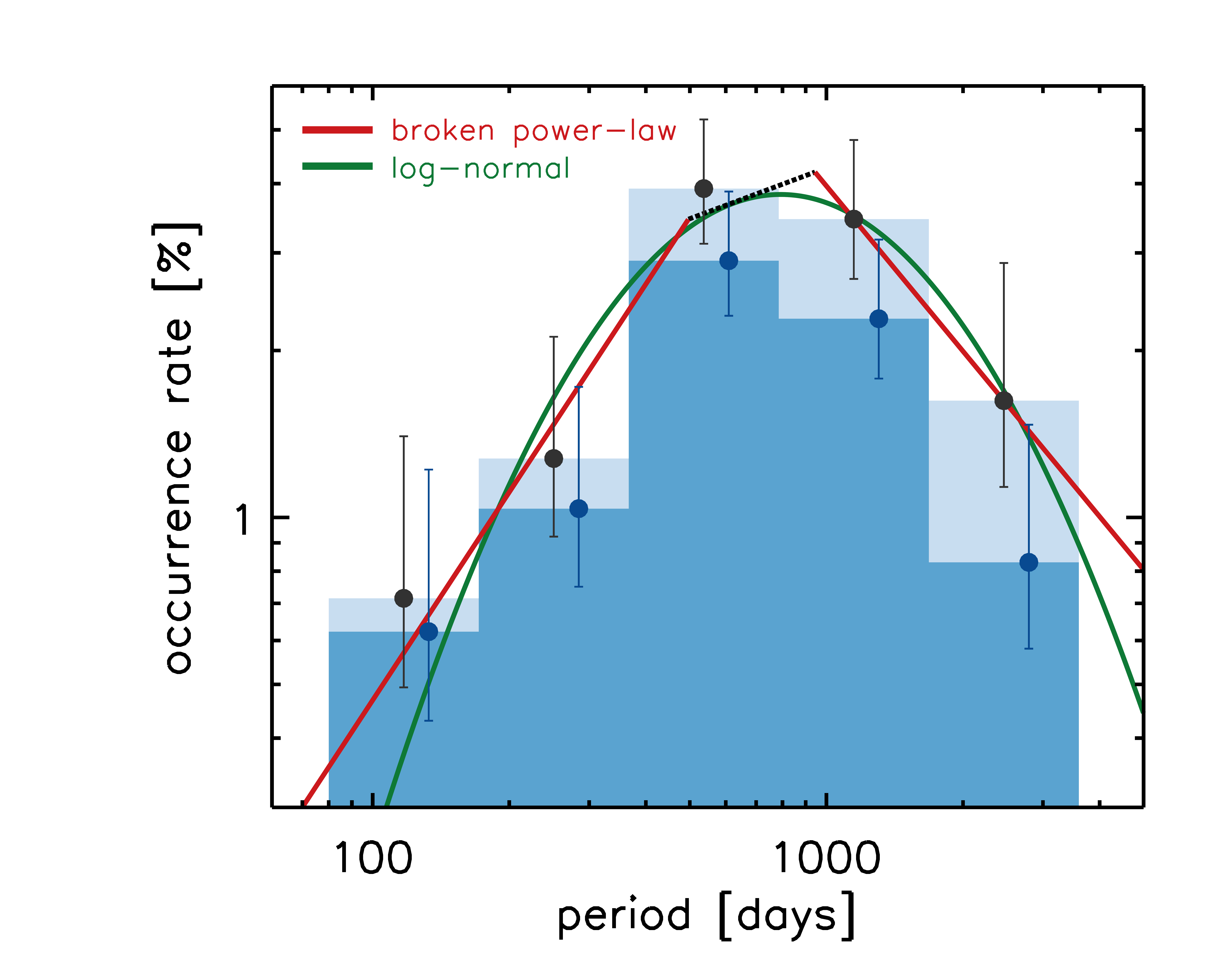}}
	\caption{Planet occurrence rate for planets with $M_\mathrm{P}\sin(i)\ge 0.8\,M_\mathrm{Jup}$ in five period bins. The blue histogram and blue dots correspond to the uncorrected rate, while the light blue histogram and black dots show the rate corrected for missed planets. Errors are derived from the binomial distribution. The log-normal fit is plotted in green, while the best-fit broken power-law is shown in red with a black dotted connector illustrating the $1\sigma$ uncertainty in $P_\mathrm{break}$.}
	\label{fig_pdist}
\end{figure}
\subsection{Giant planet period distribution around evolved stars}
To have a closer look at the period distribution, we divide the period range into five bins (which are large enough so that there are still 3 planets in the least populated bin), including the entire range of detected planet masses covered in Fig.~\ref{fig_mpplane} ($0.8\sim30\,M_\mathrm{Jup}$). We plot the resulting occurrence rates as a function of period including the completeness correction in Fig.~\ref{fig_pdist}, which clearly shows the rise and fall and the prominent peak in between, located at several hundred days. \citet{fernandes2019} find a similar behavior in a reanalysis of the giant planet occurrence rate from the HARPS and CORALIE surveys \citep{mayor2011}. Following their approach, we fit the period dependence of the occurrence rate with a broken power-law according to:
\begin{align}
	f_\mathrm{occ}\left(P\right)=C\cdot 
	\begin{cases}
		\ \left(\frac{P}{P_\mathrm{break}}\right)^{p_1}\ \mathrm{if}\ P\le P_\mathrm{break} \\[8pt]
		\ \left(\frac{P}{P_\mathrm{break}}\right)^{p_2}\ \mathrm{if}\ P> P_\mathrm{break}
	\end{cases} 
\end{align}
where $f_\mathrm{occ}(P)$ is the planet occurrence rate per period bin, $C$ is a normalization constant, $P_\mathrm{break}$ is the position of the peak in the period distribution, and $p_1$ and $p_2$ are the power-law indices of the rising and falling slope, respectively. We perform a least-squares fit and find the best-fitting parameters as: 
\begin{align*}
	C&=\left(5.48\pm 2.01\right)\\
	P_\mathrm{break}&=\left(720\pm 223\right)\,\mathrm{days}\\
	p_1&=\left(1.25\pm 0.46\right)\\
	p_2&=\left(-0.99\pm 0.81\right)
\end{align*}
\begin{figure}
	\resizebox{\hsize}{!}{\includegraphics{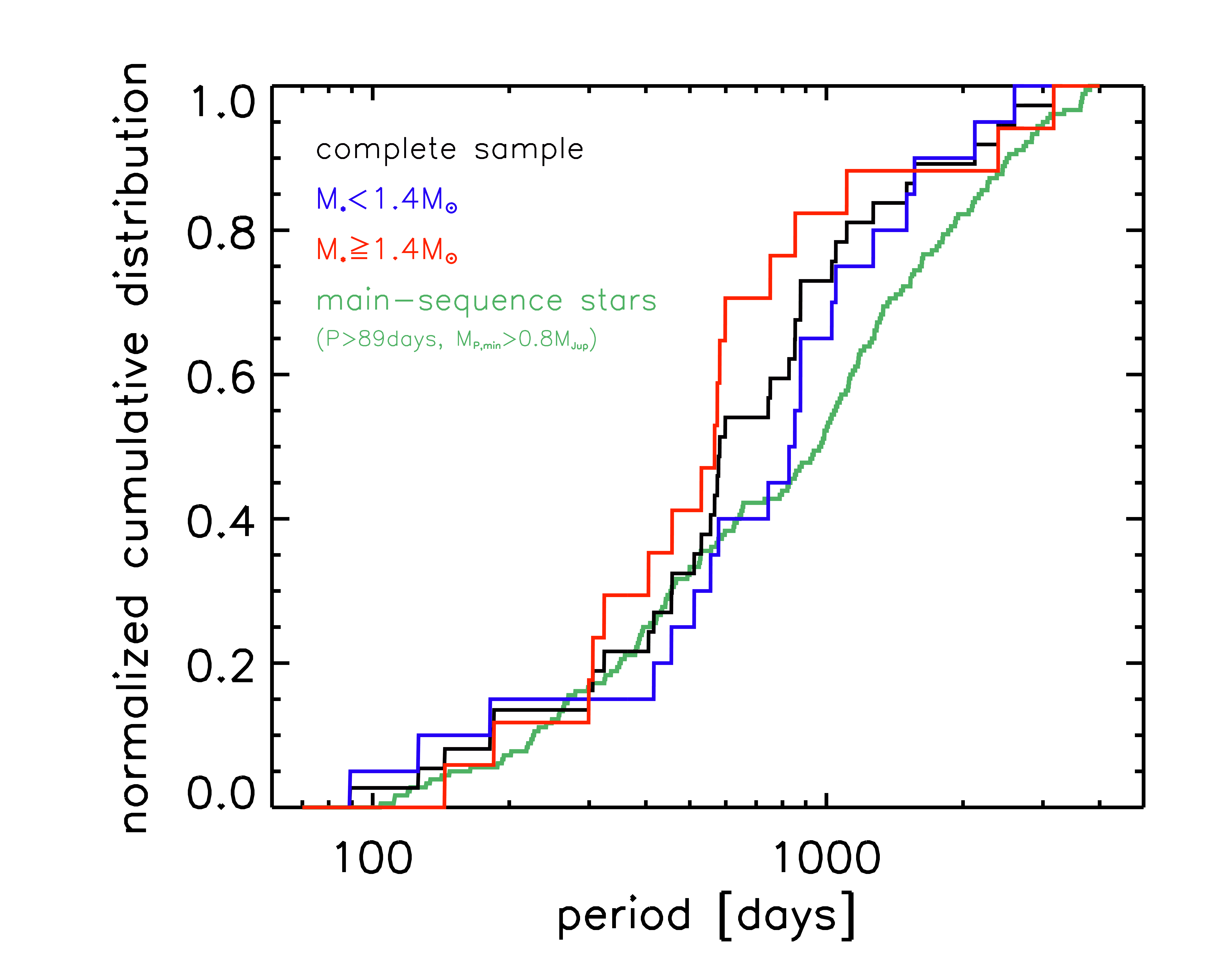}}
	\caption{Normalized cumulative distribution of orbital periods for planets with $M_\mathrm{P}\sin(i)\ge 0.8M_\mathrm{Jup}$. The black line corresponds to the entire sample, while the blue and red line show the distribution for subsamples with $M_*<1.4M_\sun$ and $M_*\ge1.4M_\sun$, respectively. For comparison, the distribution for giant planets orbiting main-sequence stars with $M_\mathrm{P}\sin(i)\ge 0.8M_\mathrm{Jup}$ and $P> 89\,\mathrm{days}$ is shown (obtained from the NASA Exoplanet Archive, selecting only planets discovered via RV and around host stars with $\log(g)\ge 4$).} 
	\label{fig_pcdf}
\end{figure}
\begin{figure}
	\resizebox{\hsize}{!}{\includegraphics{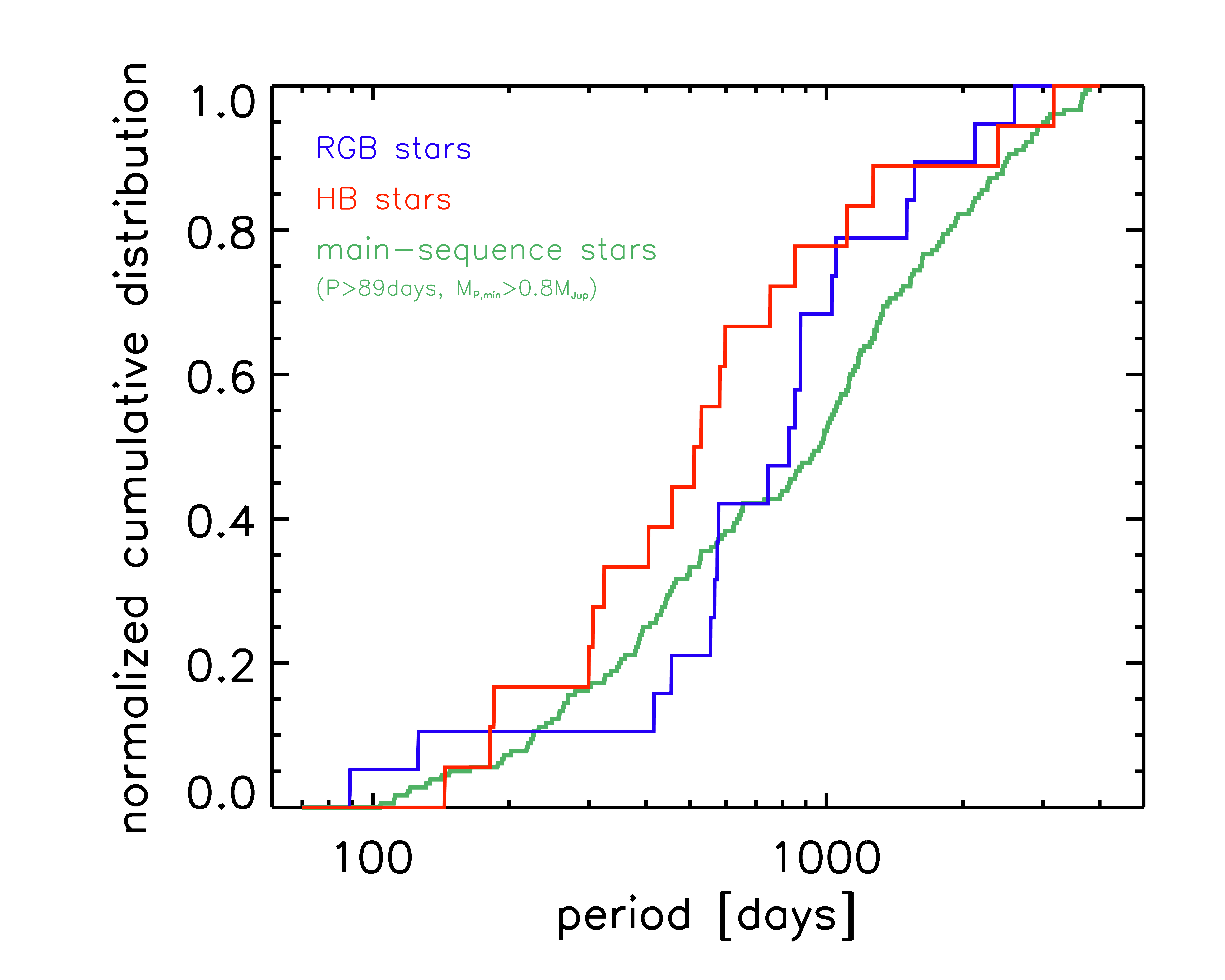}}
	\caption{Same as Fig.~\ref{fig_pcdf} but the host stars were divided into two subsets based on their evolutionary stage. The blue line shows the period distribution of planets orbiting RGB stars, while the red line corresponds to planets hosted by HB stars. The green line again traces the giant planets around main-sequence stars.}
	\label{fig_cdfevol}
\end{figure}
On a cautionary note, however, we remark that the $\chi^2$ value of 0.22 points to considerable over-fitting (as to be expected with five data points and four free parameters).\par
Additionally, we fit a log-normal distribution function: 
\begin{align}
	f_\mathrm{occ}\left(P\right)=\frac{A}{\sqrt{2\pi} \sigma \cdot P/\mathrm{days}}\cdot \exp\left(-\frac{\left(\ln(P/\mathrm{days})-\mu\right)^2}{2\sigma^2}\right) 
\end{align}
The least-squares fit yields:
\begin{align*}
	A&=\left(9988\pm 4166\right)\\
	\sigma &=\left(0.89\pm 0.19\right)\\
	\mu &=\left(7.46\pm 0.45\right)
\end{align*}
with a $\chi^2$ value of 1.14. We obtain a second estimate of $P_\mathrm{break}$ from the mode of the log-normal distribution as:
\begin{align*}
	P_\mathrm{break}[\mathrm{days}] = \exp\left( \mu - \sigma^2 \right) = 793\pm 443
\end{align*}
which is in line with the value from the broken power-law. Figure~\ref{fig_pdist} compares the result of both fits to the completeness-corrected occurrence rate in the period bins used for fitting the data.\par 
Compared to the results from \citet{fernandes2019}, who deal with main-sequence solar-mass stars, our rising slope is much steeper ($p_\mathrm{1,F19}=0.53\pm 0.09$) while the power-law index for the decrease is comparable ($p_\mathrm{2,F19}=-1.22\pm 0.47$). They find $P_\mathrm{break}$ to lie around 1200$-$2200\,days which corresponds roughly to the position of the snow line (2.3$-$3.2 AU). As our sample is on average more massive and the snow line is thought to scale approximately linearly with stellar mass, we would expect $P_\mathrm{break}$ to lie at longer periods (for a 1\,$M_\sun$ star our $P_\mathrm{break}$ corresponds to 1.6\,AU, and for a 2\,$M_\sun$ to 2\,AU). \par
To investigate if the position of the break changes with stellar mass, we divide our sample into two subsamples with $M_*<1.4$\,$M_\sun$ and $M_*\ge1.4$\,$M_\sun$. This boundary mass apportions the planets roughly evenly among the two subsamples (20 vs. 17). Unfortunately, we cannot repeat the analysis from above due to increasingly severe small number statistics leaving some period bins devoid of planets. So instead, we look at the cumulative distribution in period for both subsamples in Fig.~\ref{fig_pcdf}. Again, we can discern an increase in occurrence at around 300\,days for both subsamples (and the entire sample). In general, the overall shape of the cumulative distribution is very similar between the subsamples, and neither deviates much from the whole sample. However, there is a noticeable bump for the higher-mass subsample at $\sim$600\,days, which becomes even more apparent when we compare it to the cumulative period distribution of the RV-discovered giant planet population around main-sequence ($\log(g)\ge4$) stars (from the NASA Exoplanet Archive\footnote{\url{https://exoplanetarchive.ipac.caltech.edu/}}) -- for a better comparison, we limit the selected planets to similar planet masses and periods as found in our sample ($M_\mathrm{P}\sin(i)\ge 0.8$\,$M_\mathrm{Jup}$ and $P> 89\,\mathrm{days}$). A Kolmogorov-Smirnov (KS) test shows that the probability of our complete planet sample and the main-sequence sample being drawn from the same underlying period distribution is only 2.8\% (KS statistic $D$=0.258, $N_\mathrm{eff}$=30.7). A more detailed look at the subsamples shows this is mostly caused by the higher-mass subsample: Comparing to the main-sequence stars, the KS test finds a probability of 42\% ($D$=0.2, $N_\mathrm{eff}$=18.0) for the lower-mass subsample, and 2.5\% ($D$=0.362, $N_\mathrm{eff}$=15.5) for the higher-mass subsample, which is in line with the larger visual difference (i.e., the $\sim$600\,days bump) noted above. Both subsamples have a probability of 27\% ($D$=0.315, $N_\mathrm{eff}$=9.2) to have been drawn from the same distribution.\par
In principle, this bump could cause an enhancement in the  corresponding period bin and pull the peak of the occurrence rate to smaller periods. But in this case, the physical interpretation of the fitted $P_\mathrm{break}$ would need to change. In fact, the decrease in occurrence beyond the snow line identified by \citet{fernandes2019} is still hinted at in Fig.~\ref{fig_pcdf} as all three depicted cumulative distributions seem to change slope again after $\sim$1000\,days. As for the origin of the enhancement at $\sim$600\,days, there are several possibilities: As gas dispersal happens earlier during the formation of planets around higher-mass stars, it might be that giant planets get halted at these intermediate periods during their inward migration.\par 
On the other hand, stellar evolution can affect the orbital separation of companions and as mentioned before, there is a correlation between stellar mass and evolutionary stage in our sample due to initial target selection (see Sec.~\ref{sec_occ_global}). Therefore, we also look at two subsamples based on evolutionary stage (RGB vs HB) whose cumulative period distribution can be seen in Fig.~\ref{fig_cdfevol}. The enhancement seems to be more severe for the planets hosted by HB stars. The KS test gives a probability of 17.5\% ($D$=0.345, $N_\mathrm{eff}$=9.2) that both the RGB and the HB distributions are drawn from the same underlying distribution, while compared to the main-sequence sample, the probabilities are 24\% ($D$=0.24, $N_\mathrm{eff}$=17.2) and 6\% ($D$=0.317, $N_\mathrm{eff}$=16.4) for RGB and HB stars, respectively. And indeed one would expect that planets around HB stars have been affected more severely by stellar evolution while their host was ascending the RGB. We note however that planets with host stars above 2\,$M_\sun$ are not thought to experience significant orbital changes during the first ascent of the red-giant branch because these stars do not experience a Helium flash at the tip of the RGB and hence expand much less during their ascent, weakening the effects of both stellar tides and mass loss on the planetary orbits \citep{kunitomo2011,villaver2014}. And due to the correlation between evolutionary stage and stellar mass in our sample, we cannot fully disentangle the influence either may have on the observed period distribution.\par
A third possibility for the origin of the enhancement could be a contamination with false positives. There are several notable examples in the literature of periodic signals around giant stars that were revealed through careful observations to be most likely of stellar origin. \citet{hatzes2018} present the case of $\gamma$~Draconis showing coherent periodic RV variations with $P=702\,\mathrm{days}$ for seven years which then stopped only to pick up again after a few years, phase-shifted. Similar behavior has been detected in Aldebaran with a period of 629\,days, based on more than 500 RV measurements \citep{reichert2019}. \citet{delgadomena2018} find evidence against a planet interpretation for three giant stars with periodicities ($P=747\,\mathrm{days}$, $P=699\,\mathrm{days}$, $P=672\,\mathrm{days}$) with two of these stars showing variations in the FWHM of the cross-correlation function with a similar period, and one exhibiting a correlation between RVs and bisector inverse slope in some observing windows. All these false positives have periods comparable to where we find a possible enhancement. Oscillatory convective modes in luminous giant stars \citep[proposed in order to explain the origin of the long secondary period variables of the sequence D,][]{saio2015} have been invoked to explain the observed periods. The onset of these pulsations is associated with a minimum luminosity, the exact value however depends sensitively on the employed convection model. And considering that our HB stars are on average more luminous than our RGB stars, we cannot completely rule out a contamination with false positives.\par 
In principle, near-infrared (IR) RV data can help identify false positives as the stellar phenomena potentially mimicking planet signals are expected to exhibit different amplitudes in the IR domain -- see \citet{trifonov2015}, who obtained near-IR RVs with the CRIRES spectrograph at ESO's VLT for eight of the planet-hosting giants in our sample (from the first block of Table~\ref{table_planets}) and confirmed the consistency between the orbital solutions from optical and IR data, with specific emphasis on a stable RV semi-amplitude. Unfortunately, similar analyses are not widely available for our target class at present.
\begin{figure*}
	\centering
	\includegraphics[width=17cm]{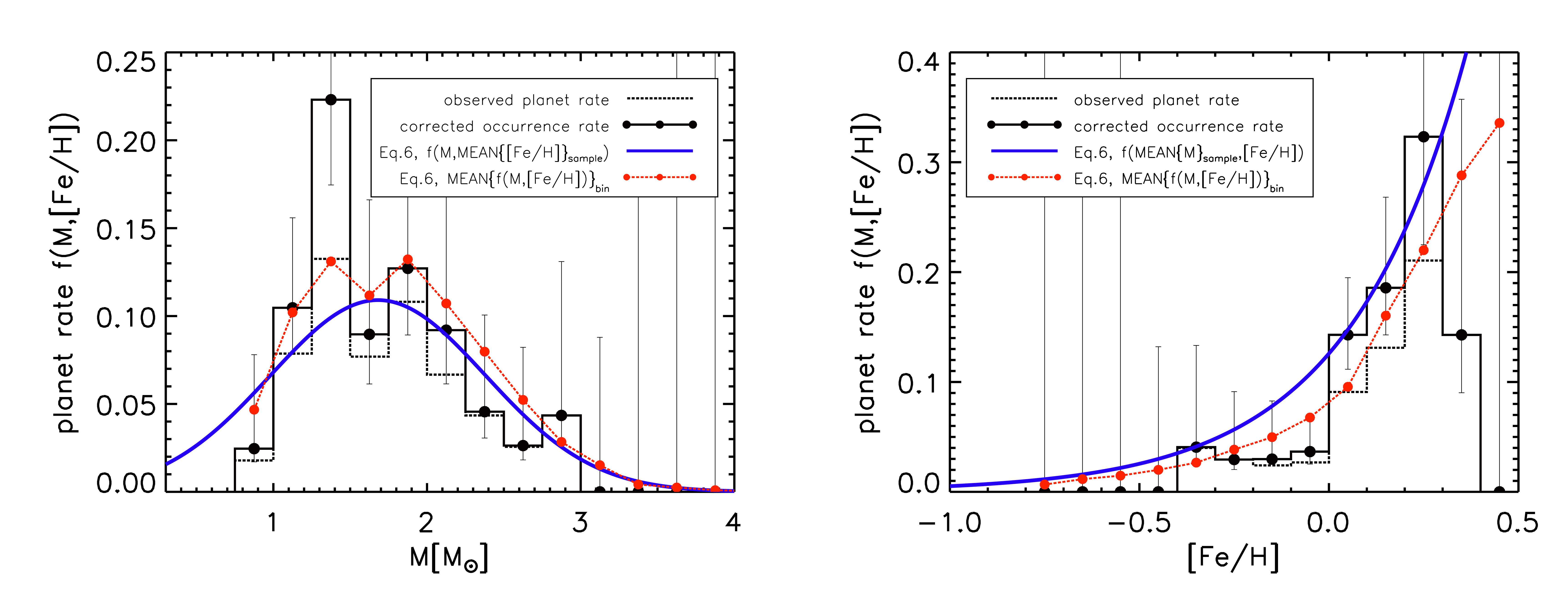}
	\caption{Planet occurrence rate as a function of stellar mass (\textit{left}) and metallicity (\textit{right}). 
		The black dotted histograms show the distribution of planets, while the solid black histograms show an estimate of the completeness correction performed for the individual bins according to the procedure outlined in Sec.~\ref{sec_occ_global}. The blue solid line represents the fitted occurrence as a function of stellar mass (\textit{left}) with the metallicity set to the sample mean of $[Fe/H]=-0.05$ and as a function of metallicity (\textit{right}) with the mass set to the sample mean of $M_*=1.81\,M_\odot$. For a more precise prediction, the red dots give the average expected occurrence rate computed using the masses and metallicities of the stars in each bin. Note that we fit the dependence on stellar mass and metallicity simultaneously and that we show the histograms only for visualization, the fit does not rely on binning of the data.}
	\label{fig_bayes}
\end{figure*}

\section{Planet occurrence as a function of stellar mass and metallicity} \label{sec_occ_mass}
In this section, we investigate the dependence of the planet occurrence rate on stellar mass and metallicity. Figure~\ref{fig_mfeh} already shows a prominent concentration of planets at higher metallicities and stellar masses between $1-2\,M_\sun$, while there are no planet hosting stars with masses larger than $3\,M_\sun$. To robustly fit the functional dependence of the occurrence rate on the stellar parameters while avoiding binning of the data and accounting for uncertainties in stellar mass and metallicity determination, we use the Bayesian inference method detailed in \citet{johnson2010}, with a small variation to include the detection probabilities derived in Sec.~\ref{sec_detprobs}: \par 
We model the data as a series of Bernoulli trials with the success probability $f$ depending on stellar mass $M_*$ and metallicity [Fe/H]. The number of trials is given by $n_\mathrm{stars}=482$ and the number of successes by the number of identified planetary systems $n_\mathrm{hosts}=32$ (hence $n_\mathrm{nonhosts}=n_\mathrm{stars}-n_\mathrm{hosts}=450$), while $d$ is the data and $X$ the set of model parameters for the functional form of the occurrence rate $f$:
\begin{equation*}
	\mathrm{P}(X|d) \propto \prod_{i=0}^{n_\mathrm{hosts}} f\left(M_{*,i}, \left[\mathrm{Fe/H}\right]_i\right) \prod_{j=0}^{n_\mathrm{nonhosts}} \left[1-f\left(M_{*,j}, \left[\mathrm{Fe/H}\right]_j\right)\cdot \overline{f_{\mathrm{c},j}}\right]
\end{equation*}
$\overline{f_{\mathrm{c},j}}$ is the mean probability of detecting a planet similar to any of the planets in the sample\footnote{As we are interested in the occurrence of planetary systems, we only consider the most easily detectable planet for multiple systems.} around star $j$ which is given by
\begin{equation*}
	\overline{f_{\mathrm{c},j}} = \frac{1}{n_\mathrm{hosts}}\cdot\sum_{i=0}^{n_\mathrm{hosts}} f_{\mathrm{R},j}\left(P_i, M_{\mathrm{P},i}\right)
\end{equation*}
where $f_{\mathrm{R},j}\left(P_i, M_{\mathrm{P},i}\right)$ is the recovery fraction of planet $i$ around star $j$ (see~Sec.~\ref{sec_detprobs}). This insertion of the mean detection probability allows for a larger $f(M_{*,j}, \left[\mathrm{Fe/H}\right]_j)$ for stars with low completeness for the detected planets and thus accounts for the fact that these stars may still host undetected planets of the same kind.\par 
\begin{figure*}
	\centering
	\includegraphics[width=17cm]{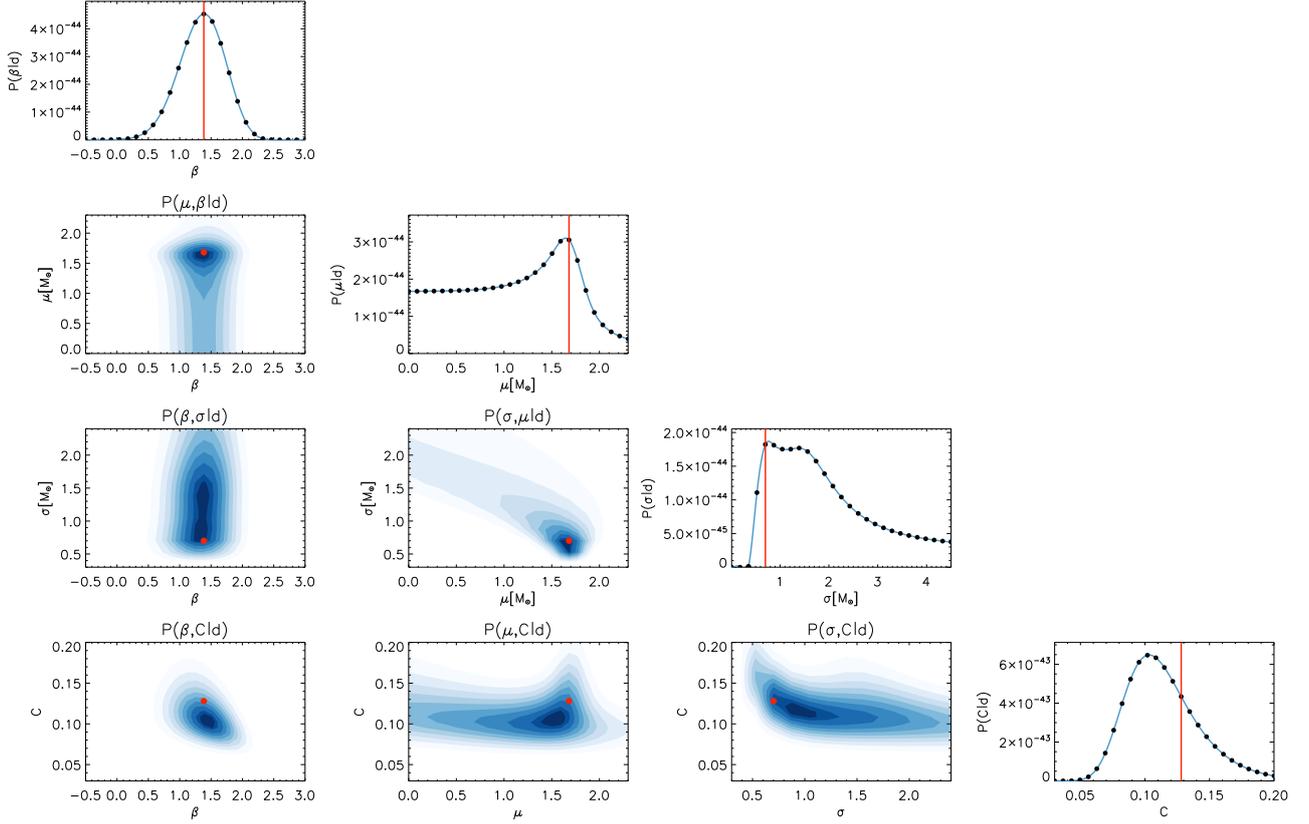}
	\caption{Marginal probability distributions functions (PDFs) from the maximum likelihood estimation for the set of model parameters from Eq.~\ref{R15}. The rightmost plot in each row shows the PDF of the indicated model parameter marginalized over the remaining three. The black dots show the grid values used for evaluating the likelihood, the blue line is a spline interpolation between them. The remaining plots show the different two-dimensional PDFs marginalized over two parameters with blue contours indicating levels from 95\% down to 10\% of the maximum probability. The red dots and vertical red lines identify the parameter values corresponding to the maximum overall likelihood. Axes are limited to the chosen uniform prior ranges.}
	\label{fig_posteriors}
\end{figure*}
The likelihood is then given by:
\begin{align*}
	\mathcal{L} = \log\left(\mathrm{P}\left(X|d\right)\right) \propto &\sum_{i=0}^{n_\mathrm{hosts}} \log f\left(M_{*,i}, \left[\mathrm{Fe/H}\right]_i\right)\\& + \sum_{j=0}^{n_\mathrm{nonhosts}} \log \left[1- f\left(M_{*,j}, \left[\mathrm{Fe/H}\right]_j\right)\cdot \overline{f_{\mathrm{c},j}}\right]
\end{align*}
We ignored $P(X)$ in the last step as it does not affect the fitting result for uniform priors on the parameters $X$. To account for uncertainties in the mass and metallicity determination, we assume a Gaussian probability distribution function centered on \{$M_{*,i}, [\mathrm{Fe/H}]_i$\} and widths corresponding to their $1\sigma$ errors and write $f\left(M_{*,i}, \left[\mathrm{Fe/H}\right]_i\right)$ as:
\begin{align*}
	f &\left( M_{*,i}, \left[\mathrm{Fe/H}\right]_i \right) = \iint f\left(M_{*},\left[\mathrm{Fe/H}\right]\right) \cdot \\ &\exp\left[ -\frac{M_*-M_{*,i}}{2\sigma_{M_{*,i}}^2} - \frac{\left[\mathrm{Fe/H}\right]-\left[\mathrm{Fe/H}\right]_{i}}{2\sigma_{\left[\mathrm{Fe/H}\right]_{i}}^2} \right] \mathrm{d}M_* \ \mathrm{d}\left[\mathrm{Fe/H}\right]
\end{align*}
where we choose the $3\sigma$ intervals as integration boundaries.
We then determine the optimal model parameters by evaluating $\mathcal{L}$ numerically over a grid and finding the set of $X$ that maximizes the likelihood.
Following \citet{reffert2015} and \citet{jones2016}, we model the dependence of the occurrence rate on stellar mass and metallicity as an exponential distribution in metallicity and a Gaussian distribution in stellar mass:
\begin{equation}
	f\left(M_*,[\mathrm{Fe/H}]\right)=C\cdot\exp\left(-\frac{1}{2}\left[ \frac{M_*-\mu}{\sigma}\right]^2\right)\cdot10^{\beta\cdot [\mathrm{Fe/H}]}\label{R15}
\end{equation}
with the model parameters $X=\{C,\mu,\sigma,\beta\}$. We find the values of maximum likelihood as 
\begin{align*}
	C &= \left(0.128\pm 0.027\right)\\
	\mu &=\left(1.68\pm 0.59\right)\,M_\sun\\
	\sigma &=\left(0.70\pm 1.02\right)\,M_\sun\\
	\beta &=\left(1.38\pm 0.36\right)
\end{align*} 
where we estimated the errors from the one-dimensional marginal probability distribution functions (PDFs; see below) via the general definition of the standard deviation for a continuous random variable. We however note that this is only a point estimator and that the PDFs themselves give a better representation of the possible parameter spaces.\par 
Figure~\ref{fig_bayes} compares the fitted model to the observed planet rates in different stellar mass and metallicity bins. The highest probability for the presence of a planet is reached at $1.68\,M_\sun$ with an occurrence rate of 12.8\% at solar metallicity.\par 
In order to show the distribution of possible values for the model parameters and potential correlations between them, we compute the one- and two-dimensional marginal PDFs according to: 
\begin{equation*}
	\mathrm{P}\left(\beta|d\right) =  \iiint \mathrm{P}\left(X|d\right)\ \mathrm{d}C\ \mathrm{d}\mu\ \mathrm{d}\sigma
\end{equation*}
and 
\begin{equation*}
	\mathrm{P}\left(\beta,\mu|d\right) =  \iint \mathrm{P}\left(X|d\right)\ \mathrm{d}C\ \mathrm{d}\sigma,
\end{equation*}
with integration limits set by the uniform priors, and analogous for the remaining parameters and their combinations. All resulting PDFs are shown in Fig.~\ref{fig_posteriors}. While there is no strong correlation between $\beta$ and any of the other parameters, the remaining three show some mutual correlations: Moving away from the maximum likelihood solution, there are regions in parameter space toward small $\mu$ and larger $\sigma$ which are not excluded by the data. This could be an artifact due to the limited mass range covered by our sample with the smallest stellar mass being $0.8\,M_\sun$ (hence in principle allowing the occurrence rate to peak outside this range if at the same time, $\sigma$ is large and hence the occurrence rate only starts dropping significantly at intermediate masses). However, other studies have shown that the occurrence is monotonically increasing at these smaller stellar masses \citep{johnson2010,ghezzi2018} and the plot for $\mathrm{P}(\mu |d)$ in Fig.~\ref{fig_posteriors} still exhibits a clear peak at the maximum likelihood value for $\mu$. Moreover, the plot for $\mathrm{P}(\sigma,\mu |d)$ shows that around the best-fit $\mu$, also $\sigma$ is fairly well constrained to below $1\,M_\sun$.\par
Additionally, we fit to our data a power-law distribution in stellar mass (Eq.~\ref{J10}, $X=\{C,\alpha,\beta\}$) like the one found by \citet{johnson2010}. As their sample was restricted to stellar masses below $\sim$$2\,M_\sun$, the model does not include a decrease in occurrence to higher masses. Hence, we include a second version (Eq.~\ref{J10exp}, $X=\{C,\alpha,M_0,\beta\}$.) of the power-law distribution with an added exponential cut-off. 
\begin{align}
	&f\left(M_*,[\mathrm{Fe/H}]\right)=C\cdot\left(\frac{M_*}{M_\odot}\right)^\alpha\cdot10^{\beta\cdot [\mathrm{Fe/H}]}\label{J10}
\end{align}
\begin{align}
	&f\left(M_*,[\mathrm{Fe/H}]\right)=C\cdot\left(\frac{M_*}{M_\odot}\right)^\alpha\cdot\exp\left(-\frac{M_*}{M_0}\right)\cdot10^{\beta\cdot [\mathrm{Fe/H}]}\label{J10exp}
\end{align}
\citet{ghezzi2018} present an update on the results of \citet{johnson2010} using the same sample and find that the occurrence rate can also be well described by a power law of the combined quantity $(M_*\times [\mathrm{Fe/H}])$, which can be considered as proportional to the amount of metals in the planet forming disk (Eq.~\ref{G18}, $X=\{C,\beta\}$.). Again, we also consider a variation that includes an exponential cut-off (Eq.~\ref{G18exp}, $X=\{C,\beta,M_0\}$.).
\begin{align}
	&f\left(M_*,[\mathrm{Fe/H}]\right)=C\cdot\left(\frac{M_*}{M_\odot}\cdot10^{[\mathrm{Fe/H}]}\right)^\beta\label{G18}
\end{align}
\begin{align}
	&f\left(M_*,[\mathrm{Fe/H}]\right)=C\cdot\left(\frac{M_*}{M_\odot}\cdot10^{[\mathrm{Fe/H}]}\right)^\beta\cdot\exp\left(-\frac{M_*}{M_0}\right)\label{G18exp}
\end{align}
Finally, we also address the notion that the planet-metallicity correlation (PMC) might be absent for evolved stars \citep[see ][]{pasquini2007,takeda2008,maldonado2013} by testing two more functional forms based on Eqs.~\ref{R15} and~\ref{J10exp} but with $\beta$ set to zero, effectively removing the dependence of the occurrence rate on metallicity:
\begin{align}
	&f\left(M_*,[\mathrm{Fe/H}]\right)=C\cdot\exp\left(-\frac{1}{2}\left[ \frac{M_*-\mu}{\sigma}\right]^2\right)\label{R15noPMC}
\end{align}
\begin{align}
	&f\left(M_*,[\mathrm{Fe/H}]\right)=C\cdot\left(\frac{M_*}{M_\odot}\right)^\alpha\cdot\exp\left(-\frac{M_*}{M_0}\right)\label{J10noPMC}
\end{align}
Table~\ref{table_params} lists the maximum likelihood model parameters for all seven tested models.
Following the procedure outlined in \citet{mortier2013}, we use the Bayes factor to compare the different functional forms and to quantify which one represents a better match to the data. It is given by 
\begin{equation}
	\mathcal{B} = \frac{\mathrm{P}\left(d|f_\mathrm{opt}\right)}{\mathrm{P}\left(d|f\right)} = \frac{\int \mathrm{P}\left(d|X_{f_\mathrm{opt}}\right)\,\mathrm{P}\left(X_{f_\mathrm{opt}}\right)\, \mathrm{d}X_{f_\mathrm{opt}}}{\int \mathrm{P}\left(d|X_{f}\right)\,\mathrm{P}\left(X_{f}\right)\, \mathrm{d}X_{f}} \text{ .}
\end{equation}

\begin{figure*}
	\centering
	\includegraphics[width=17cm]{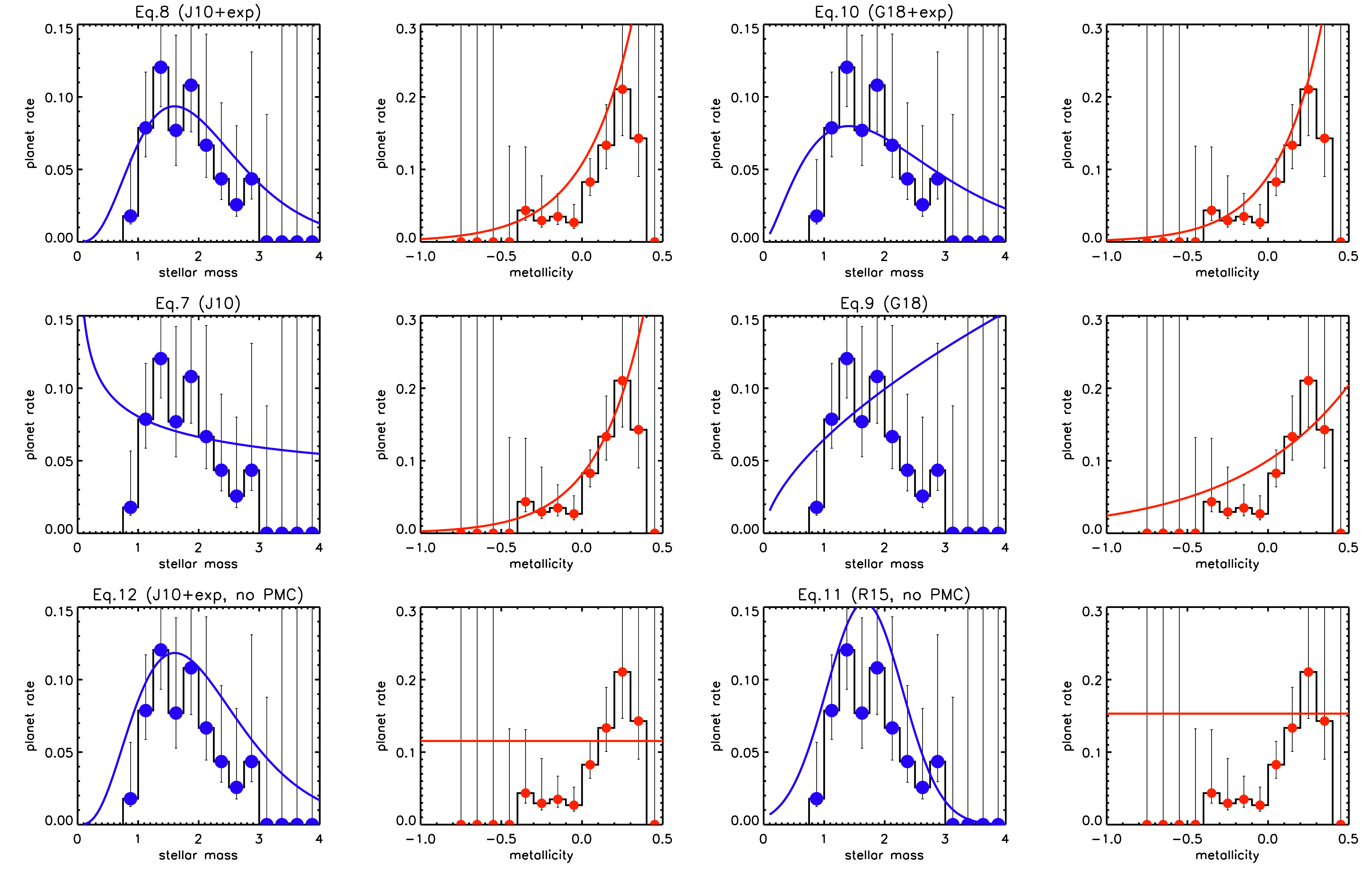}
	\caption{Similar to Fig.~\ref{fig_bayes} but for the functional forms given in Eqs.~\ref{J10}-\ref{J10noPMC}. For the mass plots, we set the metallicity to the sample mean of $-0.05$\,dex, and for the metallicity plots, we set the stellar mass to the sample mean of $1.81\,M_\odot$.}
	\label{fig_bayes_comparison}
\end{figure*}

\begin{table*}
	\caption{Fitted model parameters, incl. uncertainties, of the seven tested functional forms for the stellar mass and metallicity dependence of the planet occurrence rate and the Bayes factor for comparing between them.}
	\label{table_params}
	\centering
	{\tiny
		\begin{tabular}{l c c c c c c c}
			\hline\hline\\ [-1.7ex]
			&$C$&$\alpha$&$\beta$&$\mu$&$\sigma$&$M_0$&$\mathcal{B}$\\ [0.5ex] 
			&&&&$[M_\sun]$&$[M_\sun]$&$[M_\sun]$&\\ [0.5ex] 
			\hline \\ [-1.7ex]
			Eq.\ref{R15}: \citet{reffert2015}&0.128$\pm$0.027&$\cdots$&1.38$\pm$0.36&1.68$\pm$0.59&0.70$\pm$1.02&$\cdots$&1\\[0.3ex] 
			Eq.\ref{J10}:\citet{johnson2010}&0.096$\pm$0.025&$-$0.28$\pm$0.44&1.50$\pm$0.39&$\cdots$&$\cdots$&$\cdots$&32.5\\[0.3ex] 
			Eq.\ref{J10exp}: \citet{johnson2010} + exponential drop with stellar mass &0.673$\pm$0.167&3.41$\pm$1.19&1.45$\pm$0.41&$\cdots$&$\cdots$&0.47$\pm$0.41&2.0\\[0.3ex] 
			Eq.\ref{G18}: \citet{ghezzi2018}&0.070$\pm$0.016&$\cdots$&0.62$\pm$0.31&$\cdots$&$\cdots$&$\cdots$&5306.5\\[0.3ex] 
			Eq.\ref{G18exp}: \citet{ghezzi2018} + exponential drop with stellar mass&0.268$\pm$0.138&$\cdots$&1.57$\pm$0.39&$\cdots$&$\cdots$&0.90$\pm$0.41&6.7\\[0.3ex] 
			
			Eq.\ref{R15noPMC}: \citet{reffert2015} without PMC &0.153$\pm$0.029&$\cdots$&$\cdots$&1.67$\pm$0.56&0.63$\pm$0.99&$\cdots$&395.3\\[0.3ex]
			Eq.\ref{J10noPMC}: \citet{johnson2010} + exponential drop, without PMC&0.700$\pm$0.162&3.37$\pm$1.12&$\cdots$&$\cdots$&$\cdots$&0.48$\pm$0.39&576.4\\[0.3ex]
			\hline 
	\end{tabular}}
\end{table*}
We set Eq.~\ref{R15}, which has the highest evidence, as $f_\mathrm{opt}$ and therefore as a benchmark for comparison with the other models. The Bayes factor is also listed in Table~\ref{table_params}. Values larger than 10 indicate a strong evidence against the model, while values above 100 can be considered as decisive for model selection \citep{kass1995}. Both Eq.~\ref{J10exp} and Eq.~\ref{G18exp} show evidence comparable to our benchmark model with $\mathcal{B} < 10$ -- our data are not sufficient to decide in favor of either of the three. On the other hand, the models \ref{J10} and \ref{G18} without exponential cut-off in the stellar mass dependence have $\mathcal{B}>10$ providing strong evidence against a monotonically increasing occurrence rate with stellar mass and even decisive evidence against the functional form in Eq.~\ref{G18}. This can also be seen in Fig.~\ref{fig_bayes_comparison} that compares the best-fits for the remaining models to the data. The stellar mass dependence in Eq.~\ref{J10} even becomes negative, as for the masses present in our sample the drop in occurrence beyond the peak actually dominates over the increase at smaller masses.\par 
Overall, our results strongly support a maximum of the planet occurrence rate at intermediate stellar masses and an exponential drop thereafter. This distinct peak in the occurrence rate was already noted by \citet{reffert2015} for the Lick sample, and \citet{jones2016} for the EXPRESS sample individually, with the peak placed at $1.9^{+0.1}_{-0.5}\,M_\odot$ and $2.23^{+0.44}_{-0.06}\,M_\odot$, respectively. Based on our larger sample, which also includes the PPPS data and a completeness correction based on a detectability analysis, we find the position of the peak to lie at $1.68\,M_\odot$, which is consistent with the estimate from the Lick sample alone. The discrepancy to the value from \citet{jones2016} is mainly caused by our new parameter determination: all EXPRESS targets are now thought to have smaller masses than determined previously \citep{jones2011}; their mean stellar mass shifts from $1.86\,M_\odot$ to $1.41\,M_\odot$. \par

Regarding the PMC, the Bayes factor provides decisive evidence against the models shown in Eqs.~\ref{R15noPMC} and~\ref{J10noPMC} and hence strongly supports the presence of a dependence of planet occurrence rate on stellar metallicity. Our value for $\beta=1.38$ is smaller than found for main-sequence stars (Fischer \& Valenti 2005: $\beta=2$) but in good agreement with other results for evolved stars (\citealt{johnson2010}: $\beta=1.2\pm 0.2$; \citealt{reffert2015}: $\beta=1.7^{+0.3}_{-0.4}$; \citealt{jones2016}: $\beta=1.27^{+0.83}_{-0.42}$), though larger than the value of $\beta=1.04^{+0.12}_{-0.16}$ determined by \citet{ghezzi2018}. The question still remains as to why the PMC is weaker for evolved stars compared to what has been found for MS samples. Several contributing factors could play a role: As noted by \citet{ghezzi2018}, the occurrence of giant planets could really be a function of the total amount of solids in the proto-planetary disk. In this case, lower metallicities may be compensated for by higher stellar masses (which are not present in the MS samples). Furthermore, the PMC would also appear weaker if a fraction of our planet sample actually formed via gravitational instability, a formation channel that becomes more likely in massive disks, and which -- in contrast to core accretion -- is thought to be independent of metallicity. Moreover, it has been noted that there might be a period-metallicity correlation for giant planets \citep[see e.g., ][]{jenkins2017,petigura2018}, with metal-rich stars forming giant planets on closer orbits than metal-poor stars. During the radial expansion on the giant branch, planets around metal-rich stars would on average be more likely to get engulfed by their host star. Thereby, the PMC would appear weaker than it used to be when the stars were still on the MS. Finally, a contamination by false positives -- if present -- could also decrease the observed value of $\beta$ if it occurs independently of metallicity or is even favored by lower metallicity.\par 
We also compare our results to those of \citet{borgniet2019}, who present occurrence rates from an RV survey of A and F stars for different period, planet mass, and stellar mass bins. The bin where we can best compare our results is for $M_\mathrm{P}=1-13\,M_\mathrm{Jup}$ and $P=100-1000\,\mathrm{days}$. For stellar masses below 1.5\,$M_\sun$, they find $4.1^{+3}_{-1.2}$\% occurrence and in fact, all their detections are found in this subsample. For $M_*>1.5\,M_\sun$, they find an upper limit of $\sim$3\% (but the confidence limit goes up to $\sim$11\%). Thus, their values are smaller than our global occurrence rate but we note that their sample only encompasses few stars in the range $1.6-1.7\,M_\sun$, where we observed maximum planet occurrence. Based on our fit of the stellar mass dependence, we would expect to find about the same number of planets in bins above and below 1.5\,$M_\sun$ (assuming that there are equal numbers of stars in both bins which have a flat mass distribution), with a few more in the higher mass bin, as this basically means integrating from the peak down to either side of our occurrence rate function. Therefore, considering their large uncertainty in the higher stellar mass bin and limited completeness, as well as the fact that $\sim$\,20\% of our planets have periods larger than 1000\,days, we consider our result consistent to theirs -- so far. More observations will be necessary to unveil if the same dependence of occurrence rate on stellar mass holds for main-sequence stars, or if stellar evolution can significantly alter it in the considered period range $P>89\,\mathrm{days}$.\par 
Lastly, we revisit the question as to whether the evolutionary stage significantly affects the planet occurrence rate when factoring in the different mass distributions of the RGB and HB subsets discussed in Sec.~\ref{sec_occ_global}. For each subset, we compute the expected occurrence rate based on Eq.~\ref{R15} using the masses and metallicities of the stars in the respective subset. This yields 7.4\% and 11\% for the HB and RGB stars, respectively. Compared to the results from Sec.~\ref{sec_occ_global}, the estimate for the HB sample is in very good agreement and the result for the RGB sample lies only slightly below the lower boundary of the 1$\sigma$ error margin (11.5\%). Hence at this point, the difference in planet occurrence between our evolutionary subsamples can be almost fully explained by their different mass distributions. However, fully disentangling the effects of mass and evolutionary stage on occurrence rates would require two carefully selected evolutionary subsets with comparable mass and metallicity distributions, which is beyond the scope of this paper.

\section{Summary} \label{sec_discussion}
We presented an analysis of the combined data from the Lick, EXPRESS, and PPPS giant star RV surveys. In total, the sample consists of 482 stars hosting 37 giant planets in 32 systems, which span a range from $0.8\,M_\mathrm{Jup}$ to $24.4\,M_\mathrm{Jup}$ in minimum planet mass and from 89\,days to 3168\,days in period. The stellar parameters were homogeneously rederived using the method presented by \citet{stock2018}. We performed injection and retrieval tests of synthetic planetary signals and computed a detection probability map for each star, which we used to derive completeness corrected planet occurrence rates:
\begin{itemize}
	\item We find the global occurrence rate of planetary systems with at least one giant planet ($M_\mathrm{P,min}>0.8\,M_\mathrm{Jup}$) around the evolved stars in our sample to be $f_\mathrm{occ} = 10.7\%^{+2.2\%}_{-1.6\%}$.\\
	\item This estimate varies with evolutionary stage: for RGB stars and HB stars, we found $f_\mathrm{occ,RGB} = 14.2\%^{+4.1\%}_{-2.7\%}$ and $f_\mathrm{occ,HB} = 6.6\%^{+2.0\%}_{-1.3\%}$, respectively. However, the stellar mass distribution differs between the two subsamples and we demonstrate that this is enough to explain the observed difference in occurrence between the evolutionary subsamples.\\
	\item Our findings agree with previous studies that showed a positive planet-metallicity correlation for evolved stars \citep{wittenmyer2017a} and a distinct peak in the occurrence rate above $1.5\,M_\sun$ with an exponential drop thereafter \citep{reffert2015,jones2016}, while placing these results on a more secure statistical footing. In our Bayesian fit, maximum planet occurrence is reached at a stellar mass of $1.68\,M_\odot$, while the power-law exponent of the planet-metallicity correlation lies at $\beta=1.38$.\\
	\item The period distribution of giant planets around evolved host stars seems to peak at several hundred days. A broken power-law fit places the position of the peak at $(718\pm 226)$\,days, a log-normal fit finds ($797\pm 455$)\,days. For main-sequence stars, this break is found to occur around the snow line at 1200-2200\,days \citep{fernandes2019}. As our stars are on average more massive, the break should lie even further out. This discrepancy could be a remnant from halted migration around intermediate-mass stars, caused by stellar evolution, or an artifact from contamination by false positives in this period range. It seems to be slightly more severe for higher stellar masses and later evolutionary stages.
\end{itemize}
In conclusion, we note that for evolved stars, age can be another factor with significant influence on the occurrence and architecture of planetary systems. Moreover, the planet period and mass distributions might vary with stellar mass (or metallicity) and a simultaneous fit to all four (or even five) parameters would be ideal and worthwhile to pursue for future studies.

\begin{acknowledgements}
	V.W.\ and S.R.\ acknowledge the support of the DFG priority program SPP~1992 ``Exploring the Diversity of Extrasolar Planets'' (RE~2694/5-1). JSJ acknowledges support by FONDECYT grant 1201371, and partial support from CONICYT project Basal AFB-170002. We thank Stephan Stock for rederiving the stellar parameters of the combined sample and Paul Heeren for helpful discussions. This research has made use of the NASA Exoplanet Archive, which is operated by the California Institute of Technology, under contract with the National Aeronautics and Space Administration under the Exoplanet Exploration Program. Furthermore, this work has made use of NASA's Astrophysics Data System (ADS); of the SIMBAD database \citep{simbad} and VizieR catalog access tool \citep{vizier}, operated at CDS, Strasbourg, France; and of the TOPCAT software \citep{topcat}.
\end{acknowledgements}

\bibliographystyle{aa} 
\bibliography{bibliography}
\end{document}